%% file: main.tex
\documentclass[letterpaper, 11pt]{article}
\usepackage[utf8]{inputenc}

\usepackage{geometry}
\usepackage{amsmath,amssymb}
\usepackage{amsthm}
\usepackage{amssymb,graphicx,verbatim,enumitem,mdframed,mathrsfs}
\usepackage{mdframed}
\usepackage{subcaption}
\usepackage[toc]{appendix}

\usepackage{algorithm}
\usepackage{algorithmic}
\usepackage{arydshln}

\newtheorem{claim}{\bfseries Claim}
\newtheorem*{theorem*}{Theorem}

\newtheorem*{lemma*}{Lemma}
\newtheorem*{claim*}{Claim}

\newtheorem{theorem}{\bfseries Theorem}
\newtheorem{definition}{\bfseries Definition}
\newtheorem{corollary}{\bfseries Corollary}
\newtheorem{lemma}{\bfseries Lemma}
\newtheorem{assumption}{\bfseries Assumption}

\providecommand{\iprod}[2]{\ensuremath{\left\langle #1,\,#2  \right\rangle}}
\providecommand{\norm}[1]{\ensuremath{\left\lVert#1\right\rVert }}
\providecommand{\mnorm}[1]{\ensuremath{\left\lvert#1\right\rvert}}
\newcommand{\floor}[1]{\lfloor #1 \rfloor}

\def\R{\mathbb{R}}
\def\H{\mathcal{H}}
\def\B{\mathcal{B}}

\def\X{\mathcal{X}}

\def\V{\mathcal{V}}

\def\tm{\mathsf{Trim}}
\def\avg{\mathsf{Avg}}
\def\cge{\mathsf{CGE}}

\title{Byzantine Fault-Tolerance in \\ Peer-to-Peer Distributed Gradient-Descent}
\author{Nirupam Gupta \hspace*{.7in} Nitin H. Vaidya}
\date{\{{\it firstname}.{\it lastname}\}@georgetown.edu\\ Department of Computer Science\\ Georgetown University \\ Washington DC, USA}

\begin{document}

\maketitle

\begin{abstract}
    We consider the problem of Byzantine fault-tolerance in the peer-to-peer (P2P) distributed gradient-descent method -- a prominent algorithm for distributed optimization in a P2P system. In this problem, the system comprises of multiple agents, and each agent has a local cost function. In the fault-free case, when all the agents are honest, the P2P distributed gradient-descent method allows all the agents to reach a consensus on a solution that minimizes their aggregate cost. However, we consider a scenario where a certain number of agents may be Byzantine faulty. Such faulty agents may not follow an algorithm correctly, and may share arbitrary incorrect information to prevent other non-faulty agents from solving the optimization problem. \\
    
    In the presence of Byzantine faulty agents, a more reasonable goal is to allow all the non-faulty agents to reach a consensus on a solution that minimizes the aggregate cost of all the non-faulty agents. We formally refer to this fault-tolerance goal as {\em $f$-resilience} where $f$ is the maximum number of Byzantine faulty agents in a system of $n$ agents, with $f < n$. 
    Most prior work on fault-tolerance in P2P distributed optimization only consider {\em approximate} fault-tolerance wherein, unlike $f$-resilience, all the non-faulty agents' compute a minimum point of a {\em non-uniformly weighted} aggregate of their cost functions. We propose a fault-tolerance mechanism that confers provable $f$-resilience to the P2P distributed gradient-descent method, provided the non-faulty agents satisfy the necessary condition of {\em $2f$-redundancy}, defined later in the paper. Moreover, compared to prior work, our algorithm is applicable to a larger class of high-dimensional convex distributed optimization problems. 
\end{abstract}

\newpage
\tableofcontents

\newpage
\pagenumbering{arabic}

\section{Introduction} 
The problem of distributed optimization in a peer-to-peer (P2P) multi-agent systems has gained significant attention in recent years~\cite{boyd2011distributed, duchi2011dual, nedic2009distributed}. In this problem, each agent knows its own a {\em local} cost function. In the fault-free setting, when all the agents are non-faulty (or honest), the goal is to design a distributed algorithm that allows the agents to collectively compute a common minimum of the aggregate of their local cost functions. Specifically, we consider a peer-to-peer system architecture, as shown in Figure~\ref{fig:sys}, with $n$ agents where each agent $i$ has a local multivariate cost function $Q_i(x)$ that maps a point $x$ from the $d$-dimensional real-valued vector space $\R^d$ to a real value. Thus, for each agent $i$, $Q_i: \R^d \to \R$. In the fault-free setting, a distributed optimization algorithm allows all the agents to compute a common {\em global} minimum point
\begin{align}
    x^* \in \arg \min_{x \in \R^d} ~ \sum_{i = 1}^n Q_i(x) . \label{eqn:orig_obj}
\end{align}

As a simple example, $Q_i(x)$ may denote the cost for an agent $i$ (which may be a robot or a person) to travel to location $x$ from their current location. In this case, $x^*$ is a location that minimizes the total cost for all the agents. Such multi-agent distributed optimization 
is of interest in many practical applications,
 including distributed or {\em federated} machine learning~\cite{boyd2011distributed, kairouz2019advances}, swarm robotics~\cite{raffard2004distributed}, and large-scale sensing~\cite{rabbat2004distributed}. Most of the prior work assumes the agents to be honest (or fault-free).
 Such agents follow a specified algorithm correctly. In our work we consider a scenario wherein some of the agents may be Byzantine faulty. 
 
 \subsection{Byzantine fault-tolerance}
Su and Vaidya \cite{su2016fault} introduced the problem of
distributed optimization in the presence of a Byzantine faulty agents. A Byzantine faulty agent may behave arbitrarily~\cite{lamport1982byzantine}. In particular, the faulty agents
may share incorrect and inconsistent information in order to bias the output of a distributed optimization algorithm, and the faulty agents may also collude. For example, consider an application
of multi-agent distributed optimization to the case of sensor networks where the system has multiple sensors (or agents), and each sensor partially observes a common {\em object} in order to collectively identify the object. However, the faulty sensors may share incorrect information corresponding to arbitrary observations concocted to prevent the non-faulty agents from correctly identifying the object~\cite{chen2018resilient, chong2015observability, pajic2014robustness}. Similarly, in the case of distributed learning, which is another application of multi-agent distributed optimization, the faulty agents may send incorrect information based upon {\em mislabelled} or arbitrary concocted data points to prevent the non-faulty agents from learning a {\em good} classifier~\cite{alistarh2018byzantine, bernstein2018signsgd, blanchard2017machine, cao2019distributed, charikar2017learning, chen2017distributed, xie2018generalized}.\\

We consider a case when up to $f$ (out of $n$) of the agents are Byzantine faulty. Our goal is to design a distributed optimization algorithm that allows all the non-faulty agents to compute a minimum point of the aggregate cost function of only the non-faulty agents.
Also, we only consider {\em deterministic} algorithms which, given a fixed set of inputs from the agents, always produce the same output. We formally refer to our problem as {\em $f$-resilience}, defined as follows in the context of the aforementioned system. 

\begin{definition}[{\bf $f$-resilience}]
\label{def:t_res}
A distributed optimization algorithm $\Pi$ is said to be {\em $f$-resilient} if it allows all the non-faulty agents to compute an identical minimum point of the non-faulty agents' aggregate cost function, despite the presence of up to $f$ Byzantine faulty agents. 
\end{definition}

Alternately, an {\em $f$-resilient} distributed algorithm has the following two properties:
\begin{itemize}
    \item {\bf Consensus:} All the non-faulty agents produce the same output.
    \item {\bf Validity:} The output is a minimum point of the aggregate cost function of all the non-faulty agents.
\end{itemize}
For example, suppose that in a particular execution of an {\em $f$-resilient} distributed algorithm $\Pi$, set $\H \subseteq \{1, \ldots, \, n\}$ with $\mnorm{\H} \geq n-f$ represents the non-faulty agents. Then, each non-faulty agent $i \in \H$, implementing the algorithm $\Pi$ correctly, produces an output $\widehat{x} \in \R^d$ such that
\begin{align*}
    \widehat{x} \in \arg \min_{x \in \R^d} \sum_{i \in \H} Q_i(x).
\end{align*}
~

From prior work~\cite{gupta2020fault_podc, gupta2020resilience}, we now know that {\em $f$-resilience} is impossible unless the agents' cost functions have a certain property named {\em $2f$-redundancy}, defined below.   

\begin{definition}[{\bf $2f$-redundancy}] 
\label{def:2t_red}
The agents' cost functions are said to have {\em $2f$-redundancy} property if and only if for 
for every pair of subsets $S, \,\widehat{S} \subseteq \{1, \ldots, \, n\}$ with $\mnorm{S} = n-f$, $\mnorm{\widehat{S}} \geq n-2f$ and $\widehat{S} \subseteq S$,
$$\arg \min_{x \in \R^d} ~ \sum_{i \in \widehat{S}}Q_i(x) = \arg \min_{x \in \R^d} ~ \sum_{i \in S} Q_i(x).$$
\end{definition}


As there are at least $n-f$ non-faulty agents in the network, the \mbox{$2f$-redundancy} property implies that a point that minimizes the aggregate of any $n-2f$ non-faulty agents also minimizes the global aggregate of all the non-faulty agents, and vice-versa. The impossibility result stated below in Lemma~\ref{lem:imp} is a fundamental premise for the results presented in this paper.


\def\out{\mathsf{Out}}
\def\d{\partial}
\begin{lemma}[Theorem 1 in~\cite{gupta2020fault_podc}]
\label{lem:imp}
Suppose that each non-faulty agent's cost function is convex and differentiable. Then, there exists an {\em $f$-resilient} distributed optimization algorithm {\em only if} the non-faulty agents satisfy the {\em $2f$-redundancy} property.
\end{lemma}


We propose a fault-tolerance mechanism that robustifies the method of distributed gradient-descent - a traditional distributed optimization algorithm for peer-to-peer (P2P) systems~\cite{nedic2009distributed}. Our mechanism combines the well-known Byzantine consensus technique of coordinate-wise trimmed mean, e.g., see~\cite{su2018finite, su2016fault, sundaram2018distributed, vaidya2012matrix}, with a recently proposed {\em Byzantine robust gradient aggregation} technique named {\em comparative gradient elimination} (CGE)~\cite{gupta2019byzantine, gupta2020fault_podc}. As a key result, we show that for a bounded fraction of faulty agents $f/n$ our algorithm achieves $f$-resilience under the necessary condition of $2f$-redundancy and other standard assumptions. \\

\begin{figure}[htb!]
\centering
\includegraphics[width = 0.45\textwidth]{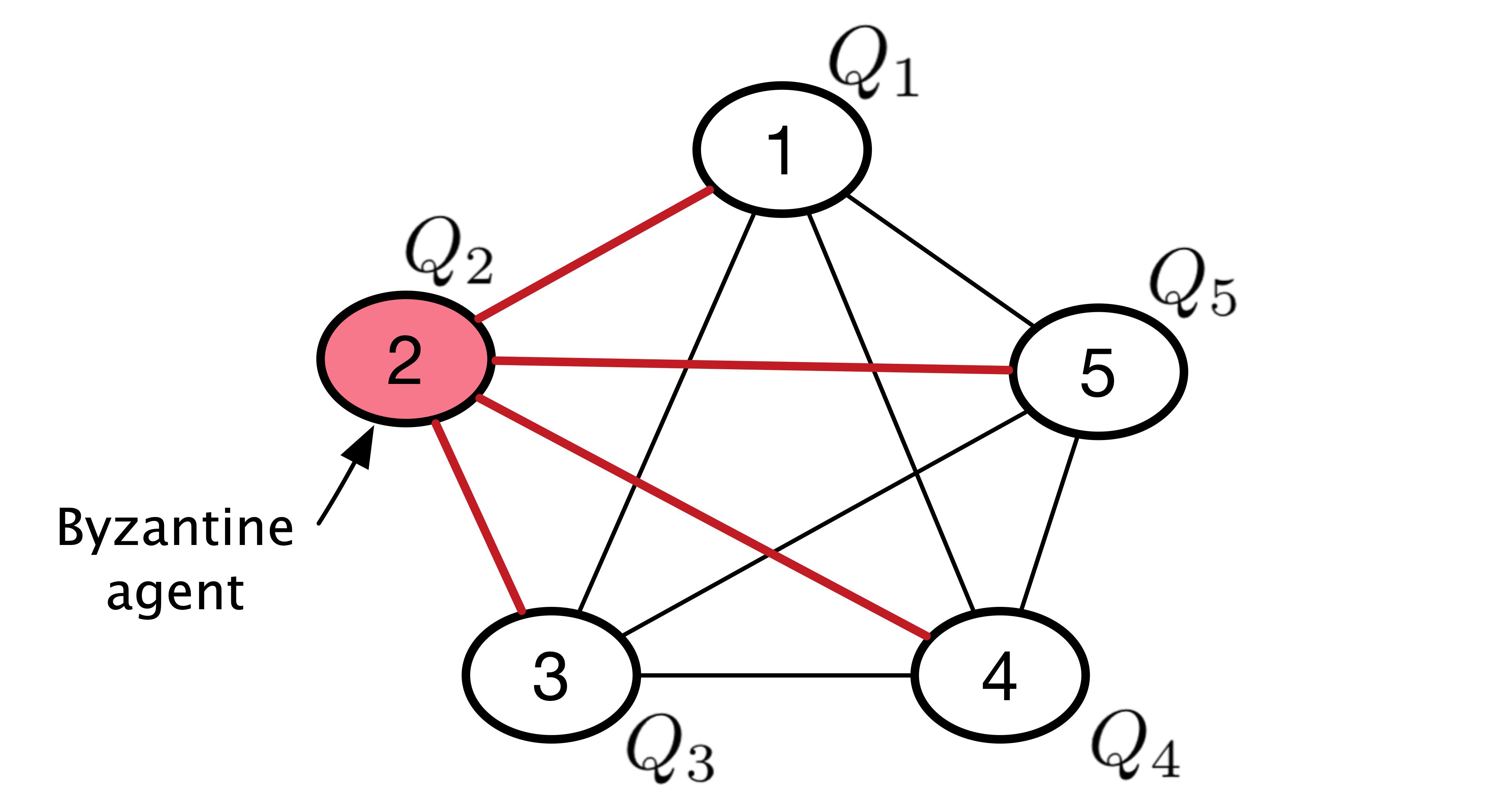}
\caption{System architecture.}
\label{fig:sys}
\end{figure}

{\bf System architecture:} The results in this paper apply to a peer-to-peer system architecture with a {\em complete} underlying communication topology, as shown in Figure~\ref{fig:sys}. The system is assumed synchronous. Unlike the server-based architecture considered in prior work such as~\cite{gupta2020fault_podc, liu2021approximate}, in peer-to-peer architecture the agents do not have access to a trustworthy server that collaborates with them to solve the optimization problem. Here, the agents must collaborate with each other, and none of the agents can be trusted by the others. Therefore, achieving $f$-resilience in a peer-to-peer architecture is more challenging than in a server-based architecture.\\

\subsection{Summary of our contributions}
\label{sub:contri}
We make the following key contributions.

\begin{itemize}
\setlength\itemsep{0.3em}
    \item Our proposed Byzantine fault-tolerance mechanism is the \underline{first-ever} to impart provable {\em $f$-resilience} to distributed gradient-descent in a P2P system.
    
    \item We show that our algorithm is {\em $f$-resilient} against a bounded fraction of faulty agents $f/n$, under the necessary condition of {\em $2f$-redundancy} and other standard assumptions. Our main result is summarized as follows.
    
    \begin{theorem*}[Informal]
    Suppose that the agents' costs have {$2f$-redundancy}, the non-faulty agents' costs are {\em $\mu$-Lipschitz} smooth, and the average of the non-faulty agents' costs is {\em $\lambda$-strongly convex}. If $f/n < (\lambda/(\lambda + 2\sqrt{d} \mu))$ then our algorithm is $f$-resilient.
    \end{theorem*}
\end{itemize}

We present below comparisons between our contributions and other related work.

\subsection{Related work}
\label{sub:prior}
The prior work on fault-tolerance in multi-agent decentralized optimization, by Su and Vaidya~\cite{su2016fault, su2020byzantine}, Sundaram and Gharesifard~\cite{sundaram2018distributed}, consider the problem of {\em approximate} Byzantine fault-tolerance wherein the agents only compute an approximate minimum point of the non-faulty aggregate cost. Specifically, they design distributed algorithms that allows the non-faulty agents to output a minimum point of some {\em non-uniformly} {weighted} aggregate of the non-faulty cost functions, instead of the uniformly weighted aggregate that we consider. More importantly, their results are only applicable to {\em scalar} cost functions, i.e., when $d = 1$. On the other hand, we consider the more general {\em high-dimensional} cost functions, i.e., when $d \geq 1$. \\

We would like to point our that the structure of our proposed algorithm is similar to that of Su and Vaidya~\cite{su2016fault, su2020byzantine} when $d = 1$. However, we use a different fault-tolerance mechanism for aggregating the agents' local gradients, namely the {\em comparative gradient-elimination} (CGE) gradient-filter as opposed to the {\em trimmed-mean} gradient-filter used in~\cite{su2016fault, su2020byzantine}. This difference is critical for achieving $f$-resilience for higher-dimensional optimization problems, i.e., when $d > 1$.\\

There is some work on {\em approximate} fault-tolerance for high-dimensional cost functions~\cite{su2018finite, su2016robust, yang2017byrdie}. However, they consider degenerate cases of the distributed optimization problem presented above. To be specific, Su and Vaidya, 2016~\cite{su2016robust} consider a setting where the individual cost functions are assumed to be linear combinations of a common set of {\em basis functions}. Yang and Bajwa, 2017~\cite{yang2017byrdie} rely on the assumption that the individual cost functions can be split into {\em independent} scalar strictly convex functions. We not make such special assumptions about the agents' cost functions, and present results for exact fault-tolerance, as opposed to approximate fault-tolerance. \\

Su and Shahrampour~\cite{su2018finite} consider the case of distributed {\em linear state estimation}, which is a special case of the more general distributed optimization problem that we consider. The fault-tolerance of their algorithm relies on an additional assumption, besides $2f$-redundancy, on the {\em data} held by the non-faulty agents. On the other hand, when applied to this special case of linear state estimation, our algorithm guarantees fault-tolerance if $2f$-redundancy holds, and the fraction of faulty agents $f/n$ is smaller than a threshold that is inversely proportional to the {\em condition number} of the non-faulty agents' {\em data matrix}.\\

In a recent work~\cite{gupta2020byzantine_thinh}, Gupta et al.~have proposed an $f$-resilient distributed optimization algorithm that is based upon the method of iterative constrained consensus~\cite{nedic2010constrained}. Their algorithm, unlike distributed gradient-descent, requires iterative projection of agents' local estimates onto their local cost functions' set of minimum points. First, in general, projection onto the set of local minimums can be computationally quite expensive. Second,

\input{algo.tex}
\section{Summary}

We have considered the problem of Byzantine fault-tolerance in the peer-to-peer (P2P) distributed gradient-descent method, which is a commonly used algorithm for solving the problem of distributed optimization in a P2P system. We have proposed a fault-tolerance mechanism that imparts exact fault-tolerance, referred as {\em $f$-resilience}, to the P2P distributed gradient-descent method in presence of up to $f$ non-faulty agents in a system with $n$ total agents. We have shown that our algorithm is $f$-resilient against a bounded fraction of faulty agents $f/n$, under the necessary condition of $2f$-redundancy and certain other standard assumptions.

\section*{Acknowledgements}
Research reported in this paper was sponsored in part by the Army Research Laboratory under Cooperative Agreement W911NF- 17-2-0196, and by the National Science Foundation award 1842198. The views and conclusions contained in this document are those of the authors and should not be interpreted as representing the official policies, either expressed or implied, of the Army Research Laboratory, National Science Foundation or the U.S. Government.

\bibliographystyle{plain}
\bibliography{ref.bib}

\newpage
\appendix

\input{app_proof_lemmas}
\input{app_proof_claims}

\end{document}

%% file: algo.tex

\section{Proposed Algorithm}
\label{sec:algo}
In this section, we first present our algorithm below in Section~\ref{sub:update}. Later, in Section~\ref{sub:ft} we present its fault-tolerance property.\\ 

Our proposed algorithm is iterative wherein each non-faulty agent maintains an estimate of a minimum of the aggregate cost of all the non-faulty agents, and updates it iteratively using local estimates and local gradients received from the other agents in the system. Specifically, for each iteration $t$, let $x^t_i$ denote the current local estimate of a non-faulty agent $i$. Each non-faulty agent $i$ sends its current local estimate $x^t_i$ and its current local gradient $\nabla Q_i(x^t_i)$ to all the other agents in the system. However, faulty agents may send different arbitrary vectors for their local estimates and gradients to different agents. Upon receiving the local estimates and gradients from all the agents, each non-faulty agent $i$ updates its current estimate $x^t_i$ using the {\em update rule} presented in equation~\eqref{eqn:update} below. \\

To be able to present the steps of our algorithm formally, we introduce below some notation. Recall that $0 \leq f < n$.\\

\subsection{Notations and Definitions}
For a vector $x \in \R^d$ its $k$-th element is denoted by $x[k]$. For two arbitrary real-valued vectors $x$ and $y$, the inner product $\iprod{x}{y}$ is defined as
\[\iprod{x}{y} = x^T y = \sum_{k = 1}^d x[k] \,y[k]\]
where $(\cdot)^T$ denotes the transpose. The Euclidean norm of a vector $x \in \R^d$ is defined to be $\norm{x} = \sqrt{\iprod{x}{x}}$. For a finite positive real value $\xi$, we let 
\[\X = [-\xi, ~ \xi]^d\]
denote a Hypercube defined as follows:
\begin{align}
    \X = \left( x \in \R^d ~ \vline ~ x[k] \in [-\xi, \, \xi], ~ \forall k \in \{1, \ldots, \, d\}\right). \label{eqn:def_hypercube}
\end{align}
The {\em Euclidean projection} of a point $x \in \R^d$ onto a closed convex set, such as $\X$, is uniquely defined to be~\cite{boyd2004convex}
\begin{align*}
    P_{\X}(x) = \arg \min_{y \in \X} \norm{x - y}.
\end{align*}
In this particular case, the $k$-th element of the projection of a point $x \in \R^d$ onto $X$, i.e., $P_{\X}(x) [k]$, for all $k \in \{1, \ldots, \, d\}$, is defined to be
\begin{align}
    P_{\X}(x) [k] = P_{[-\xi, \, \xi]}(x[k]) = \underset{y \in [-\xi, \, \xi]}{\arg \min} \norm{x[k] - y}. \label{eqn:alt_proj_X}
\end{align}
Equivalently, 
\begin{align}
    P_{\X}(x) [k] = \left\{\begin{array}{ccc} \min\left\{x[k], \, \xi \right\} & , & x[k] \geq 0 \\ ~ \\ \max \left\{ x[k], \, -\xi \right\} & , & x[k] < 0 \end{array}\right. , \quad \forall k \in \{1, \ldots, \, d\}. \label{eqn:proj_X}
\end{align}
~


{\bf Trimming ($\tm_f$):} For $n$ arbitrary (bounded) real values $a_1, \ldots, \, a_n$ such that $a_1 \leq \ldots \leq a_n$, and for $f < n/2$, we define a set
\begin{align}
    \tm_f\left\{ a_1, \ldots, \, a_n\right\} = \left\{ a_{f+1}, \ldots, \, a_{n-f}\right\}. \label{eqn:tm_f}
\end{align}
We denote the average of real-values $b_1, \ldots, \, b_n$ by $\avg\left\{b_1, \ldots, \, b_n\right\} = \frac{1}{n}\sum_{i}b_i$.\\
    
{\bf  Comparative Gradient Elimination ($\cge_f$):} For $n$ arbitrary vectors $v_1, \ldots, \, v_n$ in $\R^d$ with bounded norms and
\[\norm{v_1} \leq \ldots \leq \norm{v_n} < \infty,\]
we define a vector function 
\begin{align}
    \cge_f \left\{ v_1, \ldots, \, v_n\right\} = \sum_{i = 1}^{n-f} v_i. \label{eqn:cgc}
\end{align}

Using the above notion, we formally describe below the steps involved in each iteration of our algorithm.

\subsection{Steps in each iteration $t$}
\label{sub:update}

Before we present the steps, note that we must assume the existence of a solution in a finite search space. Otherwise, the optimization problem is rendered vacuous. Specifically, we assume the following even if the assumption is not stated explicitly again elsewhere in the paper.

\begin{assumption}[{\bf Existence}]
\label{asp:finite}
Suppose that set $\H$ with $\mnorm{\H} \geq n-f$ represents the set of all non-faulty agents. We assume that there exists $x^* \in \arg \min_{x \in \R^d} \sum_{i \in \H} Q_i(x)$ and $\xi \in (0, \, \infty)$ such that $x^* \in \X = [-\xi, \, \xi ]^d \subset \R^d$.
\end{assumption}
Recall from Definition~\eqref{eqn:def_hypercube} that set $\X$ in Assumption~\ref{asp:finite} is a $d$-dimensional Hypercube that spans from $-\xi$ to $\xi$ in each dimension. As $\xi$ can be chosen to be arbitrarily large, in practice, Assumption~\ref{asp:finite} holds true naturally~\cite{}. \\

For each pair of agents $i, \, j$ and iteration $t$, we let $m^t_{ij}$ and $g^t_{ij}$ denote the local estimate and the local gradient, respectively, received by agent $i$ from agent $j$ in iteration $t$. Note that for each \underline{non-faulty} agent $j$, for all $i$ and $t$,
\begin{align}
    m^t_{ij} = x^t_j, \text{ and } g^t_{ij} = \nabla Q_j(x^t_j). \label{eqn:shared}
\end{align}
For each non-faulty agent $j$, $m^t_{jj} = x^t_j$, and $g^t_{jj} = \nabla Q_j(x^t_j)$. If agent $j$ is \underline{faulty} then the local estimates $m^t_{ij}$ and local gradients $g^t_{ij}$ may be an arbitrary vectors, different for different $i$'s.
For an agent $i$, we let
\[\V_i = \{1, \ldots, \, n\} \setminus \{i\}.\]
For each iteration $t$ and non-faulty agent $i$, we define a vector $z^t_i \in \R^d$ such that
\begin{align}
    z^t_i[k] = \avg\left\{x^{t}_i[k], ~ \tm_f \left\{ m^t_{ij}[k], ~ j \in \V_i\right\} \right\}, \quad \forall k \in \{1, \ldots, \, d\}. \label{eqn:ctm_i}
\end{align}
Let $\X$ be a Hypercube satisfying Assumption~\ref{asp:finite} stated above. The initial local estimate for each non-faulty agent $i$, named $x^0_i$, is chosen arbitrarily from $\X$. The non-faulty agents update their local estimates iteratively by the following rule.\\

\noindent \fbox{\begin{minipage}{0.97\textwidth}
{\bf Local update rule:} In each iteration $t \in \{0, \, 1, \ldots \}$, each non-faulty agent $i$ updates its current local estimate $x^t_i$ to 
\begin{align}
    x^{t+1}_i = P_{\X}\left( z^t_i - \eta_t \, \cge_f \left\{ g^t_{i1}, \ldots, \, g^t_{in}\right\} \right) \label{eqn:update}
\end{align}
where $\eta_t$ is a positive real value referred as the {\em step-size} in iteration $t$. Projection onto $\X$ ensures boundedness of the local estimates of all non-faulty agents.
\end{minipage}}
~\\

We present below a convergence result of our algorithm.

\section{Fault-Tolerance Property}
\label{sub:ft}
In this section, we present a key result on $f$-resilient (see Definition~\ref{def:t_res}) of our algorithm presented above. Specifically, we show that under certain standard assumptions if the system has the necessary $2f$-redundancy property (defined in Definition~\ref{def:2t_red}), and the fraction of Byzantine faulty agents $f/n$ is bounded, then our algorithm satisfies the following two properties of an $f$-resilience distributed optimization algorithm. Henceforth, we let $\H \subseteq \{1, \ldots, \, n\}$ with $\mnorm{\H} \geq n-f$ denote the set of non-faulty agents. Also, recall that we assume $n > 2f$, as is needed for the necessary condition of $2f$-redundancy to hold true.

\begin{itemize}
    \item[i)] {\bf Consensus:} The local estimates of all the non-faulty agents, updating iteratively as per the update rule~\eqref{eqn:update}, eventually become equal. Specifically, 
    \begin{align}
        \lim_{t \to \infty} \norm{x_i^t - x^t_j} = 0, \quad \forall i, \, j \in \H. \label{eqn:agree}
    \end{align}
    \item[ii)] {\bf Validity:} The local estimates of each non-faulty agent $i$ converge to a minimum of the non-faulty aggregate cost function. Specifically, for all $i \in \H$, 
    \begin{align}
        \lim_{t \to \infty} x_i^t \in \arg \min_{x \in \R^d} \sum_{j \in \H} Q_j(x). \label{eqn:valid}
    \end{align}
\end{itemize}

To obtain a formal \mbox{$f$-resilience} guarantee of our algorithm we make the following standard assumptions about the non-faulty agents' cost functions~\cite{bertsekas1989parallel, bottou2018optimization, boyd2004convex}. Recall from Assumption~\ref{asp:finite} that set $\X = [-\xi, \, \xi]^d$ where $0 < \xi < \infty$.


\begin{assumption}[{\bf Lipschitz continuous gradients}]
\label{asp:lipschitz}
For each non-faulty agent $i$, we assume that the function $Q_i(x)$ is differentiable, and that the gradient $\nabla Q_i(x)$ is Lipschitz continuous in set $\X$. Specifically, for each $i \in \H$ there exists a positive finite real values $\mu$ such that 
\begin{align*}
    \norm{\nabla Q_i(x) - \nabla Q_i(y)} \leq \mu \norm{x - y}, \quad \forall x, \, y \in \X.
\end{align*}
\end{assumption}

\begin{assumption}[{\bf Strongly convex global cost}]
\label{asp:str_cvxty}
We assume that the average non-faulty cost function, defined to be
\begin{align}
    Q_{\H}(x) = \frac{1}{\mnorm{\H}} \sum_{i \in \H}Q_i(x), ~\quad \forall x,\label{eqn:def_cost_h}
\end{align}
is strongly convex in set $\X$. Specifically, there exists a positive real value $\lambda < \infty$ such that~\cite{boyd2004convex}
\begin{align*}
    \iprod{x - y}{\nabla Q_{\H}(x) - \nabla Q_{\H}(y)} \geq \lambda \norm{x - y}^2, \quad \forall x, \, y \in \X.
\end{align*}
\end{assumption}

In addition to the above assumptions on the cost functions, we also assume the step-size $\{\eta_t\}$ in the update rule~\eqref{eqn:update} to be diminishing, as stated below.

\begin{assumption}[{\bf Diminishing step-sizes}]
\label{asp:step-size}
Consider the update rule~\eqref{eqn:update} for the non-faulty agents. We assume that the sequence of step-sizes $\{\eta_t\}$ is non-increasing, i.e., $\eta_{t+1} \leq \eta_t$ for all $t \geq 0$ with $\eta_0 < \infty$, and {\em diminishing}, i.e., 
\begin{align}
    \sum_{t = 0}^\infty \eta_t = \infty \text{ and } \sum_{t = 0}^\infty \eta^2_t < \infty.
\end{align}
For example, $\eta_t = 1/(t+1)$ satisfies the above conditions~\cite{rudin1964principles}.
\end{assumption}

We note below, in Lemma~\ref{lem:bounded}, boundedness of non-faulty agents' local filtered gradients. Proof of Lemma~\ref{lem:bounded} can be found in Appendix~\ref{app:lem_bnd_cge}.

\begin{lemma}
\label{lem:bounded}
Suppose that Assumptions~\ref{asp:finite} and~\ref{asp:lipschitz} hold true. Consider the algorithm presented in Section~\ref{sub:update}. There exists a positive real value $\zeta < \infty$ such that for each non-faulty agent $i \in \H$, for all $t \geq 0$,
\begin{align}
    \norm{\cge_f \left\{ g^t_{i1}, \ldots, \, g^t_{in}\right\}} \leq \zeta . \label{eqn:bnd_cge}
\end{align}
\end{lemma}


Before we present our key result in Theorem~\ref{thm:main} below, we note in Lemma~\ref{lem:growth} below the consensus property of our algorithm. Lemma~\ref{lem:growth} relies on a prior result~\cite[Proposition 2]{su2020byzantine} on {\em consensus by iterative trimmed-mean gossip}. Also, see~\cite{anthonisse1977exponential, vaidya2012matrix}. \\

Recall that we use notation $\mnorm{\cdot}$ to denote set cardinality, absolute value of a real value and $1$-norm of a real-valued vector, depending upon its usage. Specifically, for a finite-sized set $S$, $\mnorm{S}$ denotes its cardinality, i.e., the number of elements in $S$. For $a \in \R$, $\mnorm{a}$ denotes its absolute value, and for a vector $v \in \R^d$, 
\begin{align*}
    \mnorm{v} = \sum_{k = 1}^d \mnorm{v[k]}. 
\end{align*}


\begin{lemma}
\label{lem:growth}
Suppose that Assumptions~\ref{asp:finite},~\ref{asp:lipschitz} and~\ref{asp:step-size} hold true. Consider the algorithm presented in Section~\ref{sub:update}. There exist positive real values $\rho< 1$ and $\Gamma < \infty$ such that for each pair of non-faulty agent $i, \, j \in \H$, for all $k \in \{1, \ldots, \, d\}$ and $t \geq 0$,
\begin{align}
    \mnorm{x^{t}_i[k] - x^{t}_j[k]} \leq \rho^t  \, \mnorm{\H} \, \max_{l \in \H}\mnorm{x^0_l[k]}  +  \mnorm{H} \Gamma ~ \sum_{r = 0}^{t} \eta_r \, \rho^{t - r} . \label{eqn:grwth}
\end{align}
\end{lemma}

Proof of Lemma~\ref{lem:growth} follows immediately from~\cite[Lemma 2]{su2020byzantine}. A sketch of the proof is given in Appendix~\ref{app:lem_growth}. \\

Note that under Assumption~\ref{asp:step-size}, we know that~\cite[Proposition 1]{su2016fault} 
\begin{align}
    \lim_{t \to \infty} \, \sum_{r = 0}^{t} \eta_r \, \rho^{t - r}  = 0, \quad \forall \rho \in (0, \, 1).\label{eqn:fact_sum_limit} 
\end{align}
Consequentially, Lemma~\ref{lem:growth} proves the consensus property~\eqref{eqn:agree} of our algorithm, i.e., the non-faulty agents' local estimates eventually become equal. Specifically, we obtain the following corollary of Lemma~\ref{lem:growth}.

\begin{corollary}
\label{cor:conv}
Suppose that Assumptions~\ref{asp:finite},~\ref{asp:lipschitz} and~\ref{asp:step-size} hold true. Consider the algorithm presented in Section~\ref{sub:update}. For each pair of non-faulty agent $i, \, j \in \H$,
\begin{align*}
    \lim_{t \to \infty}\norm{x^{t}_i - x^{t}_j} = 0. 
\end{align*}
\end{corollary}
~


To present our key result below in Theorem~\ref{thm:main}, we define a {\em fault-tolerance margin}
\begin{align}
    \alpha = \frac{\lambda}{\lambda + 2\sqrt{d}\, \mu} - \frac{f}{n}. \label{eqn:alpha}
\end{align}
Also, recall that under Assumption~\ref{asp:finite}, there exists $x^* \in \X = [-\xi, \, \xi]^d$ such that 
\[x^* \in \arg \min_{x \in \R^d} \sum_{i \in \H} Q_i(x).\] 
As the cost functions are assumed differentiable, 
\begin{align}
    \sum_{i \in \H} \nabla Q_i(x^*) = 0. \label{eqn:x*_zero}
\end{align}
The above, in conjunction with the strong convexity assumption, i.e., Assumption~\ref{asp:str_cvxty}, implies that $x^*$ is the only point in set $\X$ that satisfies~\eqref{eqn:x*_zero}. Consequentially, under Assumptions~\ref{asp:finite} and~\ref{asp:str_cvxty}, 
\begin{align}
    \{x^*\} = \X \cap \arg \min_{x \in \R^d} \sum_{i \in \H} Q_i(x). \label{eqn:def_x*}
\end{align}

\begin{theorem}
\label{thm:main}
Suppose that Assumptions~\ref{asp:finite},~\ref{asp:lipschitz},~\ref{asp:str_cvxty} and~\ref{asp:step-size} hold true. Consider the algorithm presented in Section~\ref{sub:update}. If the non-faulty agents $\H$ have $2f$-redundancy, and the fault-tolerance margin $\alpha > 0$, then for each agent $i \in \H$,
\begin{align*}
    \lim_{t \to \infty} \, x^t_i = x^*. 
\end{align*}
\end{theorem}

Theorem~\ref{thm:main} implies that our algorithm is provably $f$-resilient under the necessary condition of $2f$-redundancy (see Definition~\ref{def:2t_red}), if $\alpha > 0$, i.e., 
\[\frac{f}{n} < \frac{\lambda}{\lambda + 2\sqrt{d}\, \mu}.\]
We present a formal proof of Theorem~\ref{thm:main} in Section~\ref{sec:proof} below.

\section{Proof of Theorem~\ref{thm:main}}
\label{sec:proof}

In this section, we present our proof of Theorem~\ref{thm:main}. Recall that for each agent $i$, $x^t_i$ denotes its local estimate in iteration $t$. The update law for the non-faulty agents' local estimates, presented in~\eqref{eqn:update}, is as follows:
\[x^{t+1}_i = P_{\X}\left(z^t_i - \eta_t \cdot \cge_f \left\{ g^t_{i1}, \ldots, \, g^t_{in}\right\}\right), ~ \forall t \geq 0.\]
Recall from~\eqref{eqn:ctm_i} that for each $i \in \H$ and iteration $t$, 
\[z^t_i[k] = \avg\left\{x^{t}_i[k], ~ \tm_f \left\{ m^t_{ij}[k], ~ j \in \V_i\right\} \right\}, \quad \forall k \in \{1, \ldots, \, d\}.\]
For all $i$, we let
\begin{align}
    h^t_i = \cge_f \left\{ g^t_{i1}, \ldots, \, g^t_{in}\right\}, \quad \forall t. \label{eqn:not_ht}
\end{align}
Thus, for each non-faulty agent $i \in \H$,
\begin{align}
    x^{t+1}_i = P_{\X}\left(z^t_i - \eta_t \cdot h^t_i \right), ~ \forall t \geq 0. \label{eqn:update_ht}
\end{align}
Our proof relies on Claim~\ref{clm:bnd_inf_sum} below regarding the bounded infinite sum of the instantaneous differences between non-faulty agents' local estimates in each iteration. \\



\noindent \fbox{\begin{minipage}{0.97\textwidth}
\begin{claim}
\label{clm:bnd_inf_sum}
If Assumptions~\ref{asp:finite},~\ref{asp:lipschitz}  and~\ref{asp:step-size} hold true then for all $k \in \{1, \ldots, \, d\}$,
\begin{align}
    \sum_{t = 0}^{\infty} \eta_t \, \max_{i, \, j \in \H} \mnorm{x^t_i[k] - x^t_j[k]} < \infty. \label{eqn:inf_sum_diff}
\end{align}
\end{claim}
\end{minipage}}
~

\noindent (Proof of Claim~\ref{clm:bnd_inf_sum} is deferred to Appendix~\ref{app:clm_bnd}.) \\



% ...

Recall that $\H$ denotes the set of all non-faulty agents. Note that under the strong convexity assumption, i.e., Assumption~\ref{asp:str_cvxty}, the minimum point of the aggregate cost function $\sum_{i \in \H}Q_i(x)$ is unique. Recall from~\eqref{eqn:def_x*} that
\[\left\{ x^* \right\}  = \X \cap \arg \min_{x \in \R^d} \sum_{i \in \H}Q_i(x).\]
For each iteration $t$, we define a vector $x^t$ whose $k$-th element is defined to be
\begin{align}
    x^t[k] = \underset{y \in \{x^t_i[k], ~ i \in \H\}}{\arg \max} \mnorm{y - x^*[k]}, \quad \forall k \in \{1, \ldots, \, d\} .  \label{eqn:def_x_t}
\end{align}
Alternately, the $k$-th element of $x^t$ is equal to that of a non-faulty agent's current estimate whose $k$-th element is farthest from the same element of the solution $x^*$. We let $\sigma_{t,k}$ denote a non-faulty agent whose $k$-th element is equal to $x^t[k]$, i.e.,
\begin{align}
    x^t_{\sigma_{t,k}}[k] = x^t[k], \quad \forall t. \label{eqn:max_k}
\end{align}
For each $k$ and iteration $t$, we define 
\begin{align}
    V^t[k] = \left( x^t_{\sigma_{t,k}}[k] - x^*[k]\right)^2. \label{eqn:v_k}
\end{align}
Note, from~\eqref{eqn:max_k}, that for all $t$, $V^t[k] = \left(x^t[k] - x^*[k]\right)^2$, and 
\begin{align*}
    \sum_{k = 1}^d V^t[k] = \norm{x^t - x^*}^2. 
\end{align*}
We denote
\begin{align}
    V^t = \sum_{k = 1}^d V^t[k] = \norm{x^t - x^*}^2. \label{eqn:def_v_t}
\end{align}
To prove the theorem, we show that $V^t$ converges to $0$ as $t$ approaches $\infty$, i.e., 
\[\lim_{t \to \infty} V^t = 0.\]
Note that the above implies that, for each non-faulty agent $i \in \H$, $\lim_{t \to \infty}x^t_i = x^*$.

\noindent\hfil\rule{\textwidth}{.4pt}\hfil
~\\

Recall, from~\eqref{eqn:update_ht} above, that for each non-faulty agent $i$,
\[x^{t+1}_i = P_{\X}\left(z^t_i - \eta_t \cdot h^t_i \right), ~ \forall t \geq 0,\]
where $\X = [-\xi, \, \xi]^d$ is a $d$-dimensional Hypercube. Recall, from~\eqref{eqn:alt_proj_X} that for a vector $x \in \R^d$, the $k$-th element of the projected vector $P_{\X}(x)$ is defined to be
\[P_{\X}(x)[k] = P_{[-\xi, \, \xi]}\left(x[k] \right), \quad \forall k \in \{1, \ldots, \, d\}.\]
Recall from~\eqref{eqn:max_k} that $\sigma_{t, k}$ denotes a non-faulty agent whose $k$-th element is farthest from $x^*[k]$ in iteration $t$, i.e., $\sigma_{t, k} \in \H$ for all $k$ and $t$. Thus, from above, we obtain for each $k$ and $t$ that
\begin{align*}
    x^{t+1}_{\sigma_{t+1, k}}[k] = P_{[-\xi, \, \xi]}\left(z^t_{\sigma_{t+1, k}}[k] - \eta_t \, h^t_{\sigma_{t+1, k}}[k] \right).
\end{align*}
Consider an arbitrary $k \in \{1, \ldots, \, d\}$. Upon subtracting $x^*[k]$ on both sides above we obtain that 
\begin{align}
    x^{t+1}_{\sigma_{t+1, k}}[k] - x^*[k] = P_{[-\xi, \, \xi]}\left(z^t_{\sigma_{t+1, k}}[k] - \eta_t \, h^t_{\sigma_{t+1, k}}[k] \right) - x^*[k]. \label{eqn:xt_1_sigma}
\end{align}
Recall, by Assumption~\ref{asp:finite}, that $x^*[k] \in [-\xi, \, \xi]$ for all $k$ where $\xi$ is a finite positive real value. By definition of projection $P_{[-\xi, \, \xi]}(\cdot)$ in~\eqref{eqn:alt_proj_X}, we obtain that, for all $k$,
\begin{align}
    \mnorm{P_{[-\xi, \, \xi]}(x) - x^*[k]} \leq \mnorm{x[k] - x^*[k]}. \label{eqn:non-exp}
\end{align}
Recall the definition of $V^t[k]$ from~\eqref{eqn:v_k}. Using~\eqref{eqn:non-exp} in~\eqref{eqn:xt_1_sigma}, we obtain that
\begin{align}
    V^{t+1}[k] & = \left(x^{t+1}_{\sigma_{t+1, k}}[k] - x^*[k] \right)^2 \leq \left(z^t_{\sigma_{t+1, k}}[k] - \eta_t \, h^t_{\sigma_{t+1, k}}[k] - x^*[k] \right)^2 \label{eqn:exp_1}  \\
    & = \left(z^t_{\sigma_{t+1,k}}[k] - x^*[k]  \right)^2  +  \eta^2_t \left(  h^t_{\sigma_{t+1, k}}[k]\right)^2 - 2 \eta_t \, \left(z^t_{\sigma_{t+1,k}}[k] - x^*[k]  \right) \, h^t_{\sigma_{t+1, k}}[k].  \nonumber
\end{align}
Note that, as there are at most $f$ Byzantine faulty agents, 
for all $i \in \H$, $k \in \{1, \ldots, \, d\}$ and $t \geq 0$, there exists a set of non-negative real values $\{\beta^t_{i,j}[k], ~ j \in \H\}$ with $\sum_{j \in \H} \,\beta^t_{i,j}[k] = 1$ such that~\cite{vaidya2012matrix, su2020byzantine}
\begin{align}
    z^t_i[k] = \sum_{j \in \H}  \, \beta^t_{i,j}[k] \, x^t_{j}[k]. \label{eqn:cvx_z_1}
\end{align}
Thus, from~\eqref{eqn:cvx_z_1} above we note that, for all $t$ and $k$, there exists non-negative real values
$\{\beta^t_{\sigma_{t+1,k},j}[k], ~ j \in \H\}$ with $\sum_{j \in \H} \,\beta^t_{\sigma_{t+1,k},j}[k] = 1$ such that
\begin{align}
    z^t_{\sigma_{t+1,k}}[k] - x^*[k] = \sum_{j \in \H}  \beta^t_{\sigma_{t+1,k},j}[k] \left(x^t_{j}[k] - x^*[k] \right) . \label{eqn:z_t_cvx}
\end{align}
Upon squaring both sides above we obtain that
\begin{align}
    \left(z^t_{\sigma_{t+1,k}}[k] - x^*[k]  \right)^2 = \left( \sum_{j \in \H}  \beta^t_{\sigma_{t+1,k},j}[k] \left(x^t_{j}[k] - x^*[k] \right) \right)^2 \label{eqn:sub_exp_1}
\end{align}
As the square function $(\cdot)^2$ is convex~\cite{boyd2004convex}, 
\begin{align*}
    \left( \sum_{j \in \H}  \beta^t_{\sigma_{t+1,k},j}[k] \left(x^t_{\sigma_{t+1,k}}[k] - x^*[k] \right) \right)^2 \leq \sum_{j \in \H} \beta^t_{\sigma_{t+1,k},j}[k] \, \left(x^t_{j}[k] - x^*[k] \right)^2.
\end{align*}
Now, from~\eqref{eqn:max_k} note that $\mnorm{x^t_{j}[k] - x^*[k]} \leq \mnorm{x^t_{\sigma_{t, k}}[k] - x^*[k]}$ for all $j \in \H$. Upon substituting this above we obtain that
\begin{align}
    \left( \sum_{j \in \H}  \beta^t_{\sigma_{t+1,k},j}[k] \left(x^t_{\sigma_{t+1,k}}[k] - x^*[k] \right) \right)^2 & \leq \left(x^t_{\sigma_{t, k}}[k] - x^*[k] \right)^2 \, \sum_{j \in \H} \beta^t_{\sigma_{t+1,k},j}[k] \nonumber \\
    & = \left(x^t_{\sigma_{t, k}}[k] - x^*[k] \right)^2 \label{eqn:conv_above}
\end{align}
where the equality follows from the fact that $\sum_{j \in \H} \beta^t_{\sigma_{t+1,k},j}[k] = 1$. Upon substituting from~\eqref{eqn:conv_above} in~\eqref{eqn:sub_exp_1} we obtain that
\begin{align}
    \left(z^t_{\sigma_{t+1,k}}[k] - x^*[k]  \right)^2 \leq \left(x^t_{\sigma_{t, k}}[k] - x^*[k] \right)^2.\label{eqn:sub_exp_2}
\end{align}
Recall, from~\eqref{eqn:v_k}, that $\left(x^t_{\sigma_{t, k}}[k] - x^*[k] \right)^2 = V^t[k]$.
Upon substituting from~\eqref{eqn:sub_exp_2} in~\eqref{eqn:exp_1} we obtain that, for all $k$,
\begin{align*}
    V^{t+1}[k] \leq  V^t[k]  +  \eta^2_t \left(  h^t_{\sigma_{t+1, k}}[k]\right)^2 - 2 \eta_t \, \left(z^t_{\sigma_{t+1,k}}[k] - x^*[k]  \right) \, h^t_{\sigma_{t+1, k}}[k].
\end{align*}
Recall, from~\eqref{eqn:def_v_t}, that $V^t = \sum_{k = 1}^d V^t[k]$. Upon adding both sides above over all $k \in \{1, \ldots, \, d\}$, we obtain that, for all $t$,
\begin{align}
    V^{t+1} \leq V^t + \eta^2_t  \, \sum_{k = 1}^d \left(  h^t_{\sigma_{t+1, k}}[k]\right)^2 - 2 \eta_t \, \sum_{k = 1}^d \left(z^t_{\sigma_{t+1,k}}[k] - x^*[k]  \right) \, h^t_{\sigma_{t+1, k}}[k]. \label{eqn:exp_2}
\end{align}
We consider below the coefficient of $\eta_t$ in the last term on the right-hand side of~\eqref{eqn:exp_2} above, i.e., 
\[\sum_{k = 1}^d \left(z^t_{\sigma_{t+1,k}}[k] - x^*[k]  \right) \, h^t_{\sigma_{t+1, k}}[k].\]
We denote, for each $t$,
\begin{align}
    U^t = \sum_{k = 1}^d \left(z^t_{\sigma_{t+1,k}}[k] - x^*[k]  \right) \, h^t_{\sigma_{t+1, k}}[k]. \label{eqn:u_t}
\end{align}
Thus, for all $t \geq 0$,
\begin{align}
    V^{t+1} \leq V^t + \eta^2_t  \, \sum_{k = 1}^d \left(  h^t_{\sigma_{t+1, k}}[k]\right)^2 - 2 \eta_t \, U^t. \label{eqn:v_exp_3}
\end{align}
Consider an arbitrary non-faulty agent $i \in \H$. Then, we can write $U^t$ (defined in~\eqref{eqn:u_t} above) as follows, for all $t \geq 0$.
\begin{align}
    U^t = \sum_{k = 1}^d \left(x^t_i[k] - x^*[k] + z^t_{\sigma_{t+1,k}}[k] - x^t_i[k]  \right) \, h^t_{\sigma_{t+1, k}}[k]. \label{eqn:arbit_i_ut}
\end{align}
By expanding the product on the right hand side above we obtain that
\begin{align}
    U^t = \sum_{k = 1}^d \left(x^t_i[k] - x^*[k]\right) h^t_{\sigma_{t+1, k}}[k] + \sum_{k = 1}^d \left(z^t_{\sigma_{t+1,k}}[k] - x^t_i[k]  \right)  h^t_{\sigma_{t+1, k}}[k] . \label{eqn:u_t_1}
\end{align}
In~\eqref{eqn:u_t_1} above, we denote, for each $k \in \{1, \ldots, \, d\}$,
\begin{align}
    S^t[k] = \left(x^t_i[k] - x^*[k]\right) h^t_{\sigma_{t+1, k}}[k], \label{eqn:st_1}
\end{align}
and 
\begin{align}
    W^t[k] = \left(z^t_{\sigma_{t+1,k}}[k] - x^t_i[k]  \right)  h^t_{\sigma_{t+1, k}}[k]. \label{eqn:wt_1}
\end{align}
Substituting from~\eqref{eqn:st_1} and~\eqref{eqn:wt_1} in~\eqref{eqn:u_t_1} we obtain that
\begin{align}
    U^t = \sum_{k = 1}^d S^t[k] + \sum_{k = 1}^d W^t[k]. \label{eqn:u_t_2}
\end{align}
We make use of the following claims, Claim~\ref{clm:ht_ij} and~\ref{clm:zt_ij}, for obtaining a lower bound on $\sum_{k}S^t[k]$ and an upper bound on $\mnorm{\sum_{k}W^t[k]}$, respectively. Recall, from~\eqref{eqn:not_ht}, that for all $j \in \H$, 
\[h^t_j = \cge_f \left\{ g^t_{j1}, \ldots, \, g^t_{jn}\right\}.\]
~

\noindent \fbox{\begin{minipage}{0.97\textwidth}
\begin{claim}
\label{clm:ht_ij}
If the $2f$-redundancy property and Assumption~\ref{asp:lipschitz} hold true then for two arbitrary non-faulty agents $i, \, j \in \H$, for all $k \in \{1, \ldots, \, d\}$,
\begin{align}
    & \left(x^t_i[k] - x^*[k]\right) h^t_j[k] \geq  \mnorm{\H} \left(x^t_i[k] - x^*[k]\right) \nabla Q_\H(x^t_i)[k] \nonumber\\
    & - 2 \mu f \mnorm{x^t_i[k] - x^*[k]} \norm{x^t_i - x^*} - \mu \left( n + 2f \right) \, \mnorm{x^t_i[k] - x^*[k]}  \max_{l \in \H} \norm{x^t_l - x^t_i}. \label{eqn:lem_ht_j}
\end{align}
Recall, from~\eqref{eqn:def_cost_h}, that $Q_\H(x) = \frac{1}{\mnorm{\H}} \sum_{i \in \H} Q_i(x)$ for all $x$.
\end{claim}
\end{minipage}}
~

\noindent (Proof of Claim~\ref{clm:ht_ij} is deferred to Appendix~\ref{app:clm_ht}.)\\

\noindent \fbox{\begin{minipage}{0.97\textwidth}
\begin{claim}
\label{clm:zt_ij}
If Assumption~\ref{asp:lipschitz} holds true then for two arbitrary non-faulty agents $i, \, j \in \H$, for all $k \in \{1, \ldots, \, d\}$, 
\begin{align}
    \mnorm{\left(z^t_{j}[k] - x^t_i[k]  \right)  h^t_j[k]} \leq \zeta \, \max_{l \in \H} \mnorm{x^t_l[k] - x^t_i[k]} \label{eqn:clm_zt_j}
\end{align}
where recall from Lemma~\ref{lem:bounded} that $\zeta \in (0, \, \infty)$ such that $\norm{h^t_j} \leq \zeta, ~  \forall j \in \H$ and $t$.
\end{claim}
\end{minipage}}
~

\noindent (Proof of Claim~\ref{clm:zt_ij} is deferred to Appendix~\ref{app:clm_zt}.)\\

Recall the definition of $W^t[k]$ from~\eqref{eqn:wt_1} above. As a direct implication of Claim~\ref{clm:zt_ij}, we obtain that for all $k \in \{1, \ldots, \, d\}$,
\begin{align*}
    \mnorm{W^t[k]} \leq \zeta \, \max_{l \in \H} \mnorm{x^t_l[k] - x^t_i[k]}. 
\end{align*}
Therefore, 
\begin{align}
    \mnorm{\sum_{k = 1}^d W^t[k]} \leq \sum_{k = 1}^d \mnorm{W^t[k]} \leq  \zeta \, \sum_{k = 1}^d \max_{l \in \H} \mnorm{x^t_l[k] - x^t_i[k]}.\label{eqn:wt_sum_1}
\end{align}
The above implies that
\begin{align}
    \sum_{k = 1}^d W^t[k] \geq -  \zeta \, \sum_{k = 1}^d \max_{l \in \H} \mnorm{x^t_l[k] - x^t_i[k]}.\label{eqn:wt_sum_2}
\end{align}
Next, we use Claim~\ref{clm:ht_ij} to obtain a lower bound on $\sum_{k = 1}^d S^t[k]$. 

\noindent\hfil\rule{0.75\textwidth}{.4pt}\hfil
~\\

Recall from~\eqref{eqn:st_1} that, for all $k$,
\[S^t[k] = \left(x^t_i[k] - x^*[k]\right) h^t_{\sigma_{t+1, k}}[k].\]
Upon substituting from Claim~\ref{clm:ht_ij} we obtain that, for all $k$,
\begin{align*}
    S^t[k] \geq & \mnorm{\H} \left(x^t_i[k] - x^*[k]\right) \nabla Q_\H(x^t_i)[k] - 2 \mu f \mnorm{x^t_i[k] - x^*[k]} \norm{x^t_i - x^*} \nonumber \\
    & - \mu \left( n + 2f \right) \, \mnorm{x^t_i[k] - x^*[k]}  \max_{l \in \H} \norm{x^t_l - x^t_i}. 
\end{align*}
Adding both sides for all values of $k$ we obtain that
\begin{align*}
    \sum_{k = 1}^d S^t[k] \geq & \mnorm{\H} \sum_{k = 1}^d\left(x^t_i[k] - x^*[k]\right) \nabla Q_\H(x^t_i)[k] - 2 \mu f \norm{x^t_i - x^*} \, \sum_{k = 1}^d \mnorm{x^t_i[k] - x^*[k]} \nonumber \\
    & - \mu \left( n + 2f \right)  \left( \sum_{k = 1}^d\mnorm{x^t_i[k] - x^*[k]} \right) \max_{l \in \H} \norm{x^t_l - x^t_i}.
\end{align*}
Recall that the $1$-norm of a vector $v \in \R^d$, denoted by $\mnorm{v}$, and is equal to $\sum_{k = 1}^d \mnorm{v[k]}$. Substituting $\sum_{k = 1}^d\mnorm{x^t_i[k] - x^*[k]}$ by $\mnorm{x^t_i - x^*}$ above we obtain that
\begin{align*}
    \sum_{k = 1}^d S^t[k] \geq & \mnorm{\H} \sum_{k = 1}^d\left(x^t_i[k] - x^*[k]\right) \nabla Q_\H(x^t_i)[k] - 2 \mu f \norm{x^t_i - x^*} \, \mnorm{x^t_i - x^*} \nonumber \\
    & - \mu \left( n + 2f \right) \mnorm{x^t_i - x^*}\, \max_{l \in \H} \norm{x^t_l - x^t_i} . 
\end{align*}
By definition of inner product, $\sum_{k = 1}^d\left(x^t_i[k] - x^*[k]\right) \nabla Q_\H(x^t_i)[k] = \iprod{x^t_i - x^*}{Q_\H(x^t_i)}$. Substituting this above we obtain that
\begin{align}
    \sum_{k = 1}^d S^t[k] \geq & \mnorm{\H} \iprod{x^t_i - x^*}{Q_\H(x^t_i)} - 2 \mu f \norm{x^t_i - x^*} \, \mnorm{x^t_i - x^*} \nonumber \\
    & - \mu \left( n + 2f \right) \mnorm{x^t_i - x^*} \, \max_{l \in \H} \norm{x^t_l - x^t_i}  . \label{eqn:st_2}
\end{align}
Recall that for a vector $v \in \R^d$, as $\left(\sum_{k = 1}^d \mnorm{v[k]}\right)^2 \leq d \sum_{k = 1}^d \mnorm{v[k]}^2$, $\mnorm{v} \leq \sqrt{d}\norm{v}$. Using this fact in~\eqref{eqn:st_2} above we obtain that
\begin{align*}
    \sum_{k = 1}^d S^t[k] \geq & \mnorm{\H} \iprod{x^t_i - x^*}{Q_\H(x^t_i)} - 2 \sqrt{d} \mu  \, f \, \norm{x^t_i - x^*}^2 \nonumber \\
    & - \mu \sqrt{d} \left( n + 2f \right) \norm{x^t_i - x^*} \, \max_{l \in \H} \norm{x^t_l - x^t_i}.
\end{align*}
From Assumption~\ref{asp:str_cvxty}, recall that $\iprod{x^t_i - x^*}{Q_\H(x^t_i)} \geq \lambda \, \norm{x^t_i - x^*}^2$. Substituting this above we obtain that
\begin{align*}
    \sum_{k = 1}^d S^t[k] \geq \left( \lambda  \mnorm{\H}- 2 \sqrt{d} \mu f \right) \, \norm{x^t_i - x^*}^2 - \mu \sqrt{d} \left( n + 2f \right) \norm{x^t_i - x^*} \, \max_{l \in \H} \norm{x^t_l - x^t_i}.
\end{align*}
Recall that $\mnorm{\H} \geq n-f$. Thus, the above implies that
\begin{align}
    \sum_{k = 1}^d S^t[k] \geq & \left( \lambda \, n - \left( \lambda +  2 \sqrt{d} \mu \right) \, f \right) \, \norm{x^t_i - x^*}^2 \nonumber \\
    & - \mu \sqrt{d} \left( n + 2f \right) \norm{x^t_i - x^*} \,\max_{l \in \H} \norm{x^t_l - x^t_i}. \label{eqn:st_4}
\end{align}
Recall, by definition of $\alpha$ in~\eqref{eqn:alpha}, that $\lambda \, n - \left( \lambda +  2 \sqrt{d} \mu \right) \, f = n \left(\lambda + 2\sqrt{d} \mu \right) \, \alpha$. From substituting this in~\eqref{eqn:st_4} we obtain that
\begin{align}
    \sum_{k = 1}^d S^t[k] & \geq n \left(\lambda + 2\sqrt{d} \mu \right) \, \alpha \, \norm{x^t_i - x^*}^2  - \mu \sqrt{d} \left( n + 2f \right) \norm{x^t_i - x^*} \, \max_{l \in \H} \norm{x^t_l - x^t_i}. \label{eqn:st_5}
\end{align}
~

\noindent\hfil\rule{0.75\textwidth}{.4pt}\hfil
~\\

Recall from~\eqref{eqn:u_t_2} that 
\[ U^t = \sum_{k = 1}^d S^t[k] + \sum_{k = 1}^d W^t[k].\]
Substituting $\sum_{k} W^t[k]$ and $\sum_{k = 1}^d S^t[k]$ above from~\eqref{eqn:wt_sum_2} and~\eqref{eqn:st_5}, respectively, we obtain that
\begin{align}
    U^t \geq \, & n \left(\lambda + 2\sqrt{d} \mu \right) \, \alpha \, \norm{x^t_i - x^*}^2  - \mu \sqrt{d} \left( n + 2f \right) \norm{x^t_i - x^*} \, \max_{l \in \H} \norm{x^t_l - x^t_i} \nonumber \\
    & -  \zeta \, \sum_{k = 1}^d \max_{l \in \H} \mnorm{x^t_l[k] - x^t_i[k]}. \label{eqn:ut_3}
\end{align}
Now, recall from~\eqref{eqn:v_exp_3} that, for all $t \geq 0$,
\begin{align*}
    V^{t+1} \leq V^t + \eta^2_t  \, \sum_{k = 1}^d \left(  h^t_{\sigma_{t+1, k}}[k]\right)^2 - 2 \eta_t \, U^t.
\end{align*}
Substituting from~\eqref{eqn:ut_3} in the above inequality, we obtain that
\begin{align*}
    & V^{t+1} \leq V^t + \eta^2_t  \, \sum_{k = 1}^d \left(  h^t_{\sigma_{t+1, k}}[k]\right)^2 - 2 \eta_t \,n \left(\lambda + 2\sqrt{d} \mu \right) \, \alpha \, \norm{x^t_i - x^*}^2 \nonumber \\
    & + 2 \eta_t  \mu \sqrt{d} \left( n + 2f \right) \norm{x^t_i - x^*} \, \max_{l \in \H} \norm{x^t_l - x^t_i} + 2 \eta_t \zeta \, \sum_{k = 1}^d \max_{l \in \H} \mnorm{x^t_l[k] - x^t_i[k]}.
\end{align*}
We now consider a few terms on right-hand side above separately, and denote  
\begin{align}
    \Omega^t = & \eta^2_t  \, \sum_{k = 1}^d \left(  h^t_{\sigma_{t+1, k}}[k]\right)^2   + 2 \eta_t  \mu \sqrt{d} \left( n + 2f \right) \norm{x^t_i - x^*} \, \max_{l \in \H} \norm{x^t_l - x^t_i} \nonumber \\
    & + 2 \eta_t \zeta \, \sum_{k = 1}^d \max_{l \in \H} \mnorm{x^t_l[k] - x^t_i[k]}. \label{eqn:gamma_1}
\end{align}
Therefore, for all $t \geq 0$,
\begin{align}
    V^{t+1} \leq V^t - 2 \eta_t \,n \left(\lambda + 2\sqrt{d} \mu \right) \, \alpha \, \norm{x^t_i - x^*}^2 + \Omega^t. \label{eqn:vt_gamma}
\end{align}
Note that, as the fault-tolerance margin $\alpha$ is assumed positive, 
\begin{align}
     V^{t+1} \leq V^t  + \Omega^t, \quad \forall t \geq 0. \label{eqn:vt_gamma_2}
\end{align}
Recall, from the definition of $V^t$ in~\eqref{eqn:def_v_t}, that $V^t \geq 0$ for all $t$. We review below a sufficient criterion for convergence of a non-negative sequences.\\

\noindent \fbox{\begin{minipage}{0.98\textwidth}
\begin{lemma}[Bottou, 1998 \cite{bottou1998online}]
    \label{lem:seq_conv}
    Consider a sequence of real values $\{u_t, \, t = 0, \, 1, \ldots \}$. If $u_t \geq 0, \, \forall t$ then 
    \begin{align}
        \sum_{t = 0}^\infty (u_{t+1} - u_t)_{+} < \infty \implies \left\{\begin{array}{c} u_t \underset{t \to \infty}{\longrightarrow} u_\infty < \infty \\ \\ \sum_{t = 0}^\infty (u_{t+1} - u_t)_{-} > -\infty \end{array}\right.
    \end{align}
    where the operators $(\cdot)_{+}$ and $(\cdot)_{-}$ are defined as follows for a real scalar $x$,
    \begin{align*}
        (x)_{+} = \left\{\begin{array}{ccc} x &, & x > 0\\ 0 &, & \text{otherwise} \end{array}\right. \text{, and } (x)_{-} = \left\{\begin{array}{ccc} 0 &, & x > 0\\ x &, & \text{otherwise} \end{array}\right.
    \end{align*}
\end{lemma}
\end{minipage}}
~\\

We show below that $\sum_{t = 0}^\infty \left(V^{t+1} - V^t\right)_{+} < \infty$. This, in conjunction with Lemma~\ref{lem:seq_conv}, will prove that $\lim_{t \to \infty} V^t < \infty$. \\

Note, from~\eqref{eqn:vt_gamma_2}, that for all $t \geq 0$,
\begin{align*}
    \left(V^{t+1} - V^t\right)_{+} \leq \Omega^t
\end{align*}
where observer from~\eqref{eqn:gamma_1} that $\Omega^t \geq 0$.
Therefore, 
\begin{align}
    \sum_{t = 0}^\infty \left(V^{t+1} - V^t\right)_{+} \leq \sum_{t = 0}^\infty \Omega^t. \label{eqn:vt_gamma_3}
\end{align}
Recall from~\eqref{eqn:gamma_1} that, for all $t$,  
\begin{align}
    \Omega^t = & \eta^2_t  \, \sum_{k = 1}^d \left(  h^t_{\sigma_{t+1, k}}[k]\right)^2   + 2 \eta_t  \mu \sqrt{d} \left( n + 2f \right) \norm{x^t_i - x^*} \,\max_{l \in \H} \norm{x^t_l - x^t_i} \nonumber \\
    & + 2 \eta_t \zeta \, \sum_{k = 1}^d \max_{l \in \H} \mnorm{x^t_l[k] - x^t_i[k]}. \label{eqn:om_ht_k}
\end{align}
Recall that for a vector $v \in \R^d$, $\mnorm{v[k]} \leq \norm{v}$ for each $k \in \{1, \ldots, \, d\}$. Therefore, in~\eqref{eqn:om_ht_k}, $\left(  h^t_{\sigma_{t+1, k}}[k]\right)^2 \leq \norm{h^t_{\sigma_{t+1, k}}}^2$, and we obtain that
\begin{align}
    \Omega^t \leq & \eta^2_t  \, \sum_{k = 1}^d \norm{h^t_{\sigma_{t+1, k}}}^2  + 2 \eta_t  \mu \sqrt{d} \left( n + 2f \right) \norm{x^t_i - x^*} \, \max_{l \in \H} \norm{x^t_l - x^t_i}\nonumber \\
    & + 2 \eta_t \zeta \, \sum_{k = 1}^d \max_{l \in \H} \mnorm{x^t_l[k] - x^t_i[k]}. \label{eqn:gamma_2}
\end{align}
Recall, from~\eqref{eqn:not_ht}, that 
\[h^t_{\sigma_{t+1, k}} = \cge_f \left\{ g^t_{\sigma_{t+1, k} \, 1}, \ldots, \, g^t_{\sigma_{t+1, k} \, n} \right\}.\]
As $\sigma_{t+1, k} \in \H$, Lemma~\ref{lem:bounded} implies that $\norm{h^t_{\sigma_{t+1, k}}} \leq \zeta$, for all $k$ and $t$, where $\zeta < \infty$. Therefore, $\sum_{k = 1}^d \left(  h^t_{\sigma_{t+1, k}}[k]\right)^2 \leq d \, \zeta^2$. Substituting this in~\eqref{eqn:gamma_2} implies that
\begin{align}
    \Omega^t \leq & \eta^2_t  \, d \, \zeta^2  + 2 \eta_t  \mu \sqrt{d} \left( n + 2f \right) \norm{x^t_i - x^*} \, \max_{l \in \H} \norm{x^t_l - x^t_i} + 2 \eta_t \zeta \, \sum_{k = 1}^d \max_{l \in \H} \mnorm{x^t_l[k] - x^t_i[k]}. \label{eqn:gamma_3}
\end{align}
Now, owing to triangle inequality, $\norm{x^t_i - x^*} \leq \norm{x^t_i} + \norm{x^*}$. As agent $i \in \H$, recall from~\eqref{eqn:update}, that $x^t_i \in \X = [-\xi, \, \xi]^d$ for all $t$. Also, recall that $x^* \in \X$. Thus, 
\[\norm{x^t_i - x^*} \leq \norm{x^t_i} + \norm{x^*} \leq  d \xi + d \xi = 2 \, d \xi.\] 
Substituting from above in~\eqref{eqn:gamma_3} we obtain that
\begin{align}
    \Omega^t \leq & \eta^2_t  \, d \, \zeta^2  + 4 \eta_t  \mu d \sqrt{d} \xi \left( n + 2f \right) \, \max_{l \in \H} \norm{x^t_l - x^t_i} + 2 \eta_t \zeta \, \sum_{k = 1}^d \max_{l \in \H} \mnorm{x^t_l[k] - x^t_i[k]}. \label{eqn:gamma_4}
\end{align}
For vector $v \in \R^d$, $\norm{v} \leq \sum_{k = 1}^d \mnorm{v[k]}$. Therefore, for all $i, \, l \in \H$, $\norm{x^t_l - x^t_i} \leq \sum_{k = 1}^d \mnorm{x^t_l[k] - x^t_i[k]}$. This implies that $\max_{l \in \H} \norm{x^t_l - x^t_i} \leq \sum_{k = 1}^d \max_{l \in \H} \mnorm{x^t_l[k] - x^t_i[k]}$. Upon using this inequality in~\eqref{eqn:gamma_4} we obtain that
\begin{align*}
    \Omega^t \leq & \eta^2_t  \, d \, \zeta^2  + 2 \eta_t  \left(2 \mu d \sqrt{d} \xi \left( n + 2f \right) + \zeta \right)\, \sum_{k = 1}^d \max_{l \in \H} \mnorm{x^t_l[k] - x^t_i[k]}.
\end{align*}
Note that above holds true for each iteration $t \in \{0, \, 1, \ldots \}$. Therefore,
\begin{align}
    \sum_{t = 0}^\infty \Omega^t \leq & d \, \zeta^2 \sum_{t = 0}^\infty \eta^2_t + 2 \left(2 \mu d \sqrt{d} \xi \left( n + 2f \right) + \zeta \right) \sum_{t = 0}^\infty \eta_t \, \sum_{k = 1}^d \max_{l \in \H} \mnorm{x^t_l[k] - x^t_i[k]}. \label{eqn:gamma_5} 
\end{align}
By Assumption~\ref{asp:step-size}, $\sum_{t = 0}^\infty \eta^2_t < \infty$. Thus, as $d$ and $\zeta$ are finite constants, 
\begin{align}
    d \zeta^2 \sum_{t = 0}^\infty \eta^2_t  < \infty. \label{eqn:gamma_5_1}
\end{align}
From Claim~\ref{clm:bnd_inf_sum} we obtain that, under Assumptions~\ref{asp:finite},~\ref{asp:lipschitz} and~\ref{asp:step-size},
\begin{align}
    \sum_{t = 0}^\infty \eta_t \sum_{k = 1}^d \max_{l \in \H} \mnorm{x^t_l[k] - x^t_i[k]} =  \sum_{k = 1}^d \sum_{t = 0}^\infty \eta_t  \max_{l \in \H} \mnorm{x^t_l[k] - x^t_i[k]} < \infty. \label{eqn:gamma_5_2}
\end{align}
As $\xi < \infty$ (Assumption~\ref{asp:finite}), substituting from~\eqref{eqn:gamma_5_1} and~\eqref{eqn:gamma_5_2} in~\eqref{eqn:gamma_5} we obtain that
\begin{align}
    \sum_{t = 0}^\infty \Omega^t < \infty. \label{eqn:gamma_finite}
\end{align}
Substituting from~\eqref{eqn:gamma_finite} in~\eqref{eqn:vt_gamma_3}, we obtain that
\begin{align}
    \sum_{t = 0}^\infty \left(V^{t+1} - V^t\right)_{+} \leq \sum_{t = 0}^\infty \Omega^t < \infty.\label{eqn:v_inf_v_0} 
\end{align}
As $V^t \geq 0$ for all $t$ (see~\eqref{eqn:def_v_t}), the above, in conjunction with Lemma~\ref{lem:seq_conv} implies that
\begin{align}
    V^\infty = \lim_{t \to \infty} V^t < \infty. \label{eqn:v_conv}
\end{align}
~

Now, recall from~\eqref{eqn:vt_gamma} that for all $t \geq 0$,
\begin{align*}
    V^{t+1} \leq V^t - 2 \eta_t \,n \left(\lambda + 2\sqrt{d} \mu \right) \, \alpha \, \norm{x^t_i - x^*}^2 + \Omega^t. 
\end{align*}
Therefore, 
\begin{align*}
    \sum_{t = 0}^\infty V^{t+1} \leq \sum_{t = 0}^\infty V^t - 2 n \left(\lambda + 2\sqrt{d} \mu \right)  \alpha \, \sum_{t = 0}^\infty \eta_t \, \norm{x^t_i - x^*}^2 + \sum_{t = 0}^\infty \Omega^t. 
\end{align*}
Note that for all $\tau \in \{0, \, 1, \ldots \}$, $\sum_{t = 0}^\tau V^{t+1} - \sum_{t = 0}^\tau V^t = V^{\tau + 1} - V^0$. Thus, from above we obtain that
\begin{align}
    V^{\infty} - V^0 \leq - 2 n \left(\lambda + 2\sqrt{d} \mu \right)  \alpha \, \sum_{t = 0}^\infty \eta_t \, \norm{x^t_i - x^*}^2 + \sum_{t = 0}^\infty \Omega^t.  \label{eqn:sum_vt_gamma}
\end{align}
From~\eqref{eqn:gamma_finite}, recall that $\sum_{t = 0}^\infty \Omega^t < \infty$. From~\eqref{eqn:v_conv}, recall that $V^\infty < \infty$.
As per the specifications of the algorithm, the initial local estimate $x^0_i$ of each non-faulty agent has bounded norm, i.e, $\norm{x^0_i} < \infty$, and by Assumption~\ref{asp:finite}, $\norm{x^*} < \infty$. Therefore, by definition of $V^t$ in~\eqref{eqn:def_v_t}, $V^0 < \infty$. Thus, from~\eqref{eqn:sum_vt_gamma} we obtain that
\begin{align*}
    \mnorm{2 n \left(\lambda + 2\sqrt{d} \mu \right)  \alpha \, \sum_{t = 0}^\infty \eta_t \, \norm{x^t_i - x^*}^2} < \infty.
\end{align*}
As $\alpha$ is assumed positive and is bounded by Definition~\eqref{eqn:alpha}, the above implies that
\begin{align}
    \sum_{t = 0}^\infty \eta_t \, \norm{x^t_i - x^*}^2 < \infty. \label{eqn:sum_et_xti}
\end{align}
Recall that $i$ in~\eqref{eqn:sum_et_xti} is an arbitrary non-faulty agent in $\H$. Therefore,~\eqref{eqn:sum_et_xti} holds true for all $i \in \H$. Therefore, from~\eqref{eqn:sum_et_xti} we obtain that
\begin{align}
    \sum_{i \in \H} \sum_{t = 0}^\infty \eta_t \, \norm{x^t_i - x^*}^2 < \infty. \label{eqn:sum_et_xti_1}
\end{align}
Obviously, from the definition of $V^t$ in~\eqref{eqn:def_v_t}, for all $t$,
\begin{align*}
    V^t \leq \sum_{i \in \H} \norm{x^t_i - x^*}^2.
\end{align*}
As $\eta_t \geq 0$ for all $t$, the above implies that
\begin{align}
    \eta_t \, V^t \leq \sum_{i \in \H} \eta_t \, \norm{x^t_i - x^*}^2, \quad \forall t. \label{eqn:sum_et_xti_2}
\end{align}
From~\eqref{eqn:sum_et_xti_1} and~\eqref{eqn:sum_et_xti_2} we obtain that
\begin{align}
    \sum_{t = 0}^\infty \eta_t V^t < \infty. \label{eqn:sum_et_xti_3}
\end{align}
Recall that, by Assumption~\ref{asp:step-size}, $\sum_{t = 0}^\infty \eta_t = \infty$. Finally, we reason by contradiction to show that $V^\infty = \lim_{t \to \infty} V^t = 0$.\\

Suppose, toward a contradiction, that $V^\infty = \delta > 0$. Therefore, as the sequence $\{V^t, ~ t = 0, \, 1, \ldots \}$ converges to $V^\infty$, there exists a positive integer $\tau < \infty$ such that 
\begin{align*}
    V^t > V^\infty - \frac{\delta}{2} = \frac{\delta}{2} ~ , \quad \forall t > \tau.
\end{align*}
The above implies that
\begin{align}
    \sum_{t = \tau+1}^\infty \eta_t V^t > \frac{\delta}{2} \, \sum_{t = \tau+1}^\infty \eta_t. \label{eqn:contra_1}
\end{align}
By Assumption~\ref{asp:step-size}, $ \sum_{t = \tau}^\infty \eta_t = \infty$. Thus,~\eqref{eqn:contra_1} implies that
\begin{align}
    \sum_{t = \tau + 1}^\infty \eta_t \, V^t = \infty. \label{eqn:contra_2}
\end{align}
As both $\eta_t$ and $V^t$ are non-negative for all $t \geq 0$,~\eqref{eqn:contra_2} implies that
\begin{align}
    \sum_{t = 0}^\infty \eta_t \, V^t \geq \sum_{t = \tau + 1}^\infty \eta_t \, V^t = \infty. \label{eqn:contra_3}
\end{align}
The above, i.e.,~\eqref{eqn:contra_3}, contradicts the proven fact~\eqref{eqn:sum_et_xti_3}. Therefore, 
$$V^\infty = \lim_{t \to \infty} V^t = 0.$$
Hence, by definition of $V^t$ in~\eqref{eqn:def_v_t} and $x^t$ in~\eqref{eqn:def_x_t}, 
\begin{align*}
    \lim_{t \to \infty} x^t_i = x^*, \quad \forall i \in \H.
\end{align*}

%% file: app_proof_lemmas.tex
\section{Appendix: Proof of Lemma~\ref{lem:bounded}}
\label{app:lem_bnd_cge}

\noindent \fbox{\begin{minipage}{0.97\textwidth}
\begin{lemma*}[Restated]
Suppose that Assumptions~\ref{asp:finite} and~\ref{asp:lipschitz} hold true. Consider the algorithm presented in Section~\ref{sub:update}. There exists a positive real value $\zeta < \infty$ such that for each non-faulty agent $i \in \H$, for all $t \geq 0$,
\begin{align*}
    \norm{\cge_f \left\{ g^t_{i1}, \ldots, \, g^t_{in}\right\}} \leq \zeta .
\end{align*}
\end{lemma*}
\end{minipage}}
~

\begin{proof}
Consider an arbitrary non-faulty agent $i \in \H$ and an iteration $t$. Recall that $g^t_{ij}$ denotes the gradient send by agent $j$ to agent $i$ in iteration $t$. Also, recall that if $j \in \H$ then $g^t_{ij} = \nabla Q_j(x^t_j)$. A faulty agent $j$ may send an arbitrary vector for its gradient $g^t_{ij}$. We denote by $i_j \in \{1, \ldots, \, n\}$ the agent that sends gradient with $j$-th smallest Euclidean norm. Specifically,  
\begin{align}
    \norm{g^t_{i \, i_1}} \leq \ldots \leq \norm{g^t_{i \, i_{n-f}}} \leq \ldots \leq \norm{g^t_{i \, i_{n}}}. \label{eqn:order_norm_i}
\end{align}
From~\eqref{eqn:order_norm_i}, and the definition of $\cge_f$ in~\eqref{eqn:cgc}, we obtain that
\begin{align}
    \norm{\cge_f \left\{ g^t_{i1}, \ldots, \, g^t_{in}\right\}} \leq (n-f) \norm{g^t_{i \, i_j}}, \quad \forall j \in \{n-f, \ldots, \, n \}.  \label{eqn:cge_order_i}
\end{align}
As there are at most $f$ Byzantine faulty agents, there exits a non-faulty agent $\nu_i \in \{i_{n-f}, \ldots, \, i_n\}$. Recall that, as $\nu_i \in \H$, $g^t_{i \, \nu_i} = \nabla Q_{\nu_i} (x^t_{\nu_i})$. Thus, from~\eqref{eqn:cge_order_i} we obtain that
\begin{align}
    \norm{\cge_f \left\{ g^t_{i1}, \ldots, \, g^t_{in}\right\}} \leq (n-f) \norm{g^t_{i \, \nu_i}} = (n-f) \norm{\nabla Q_{\nu_i} (x^t_{\nu_i})}.  \label{eqn:cge_nu_i}
\end{align}
Recall, from Assumption~\ref{asp:lipschitz}, that the cost function of each non-faulty $i \in H$ is differentiable. Therefore, for all $i \in \H$, 
\begin{align}
    \text{if } \norm{x} < \infty \text{ then } \norm{\nabla Q_i(x)} < \infty. \label{eqn:grad_exists}
\end{align}
Now, recall from the update rule~\eqref{eqn:update} that, for all $i \in \H$ and iteration $t \geq 0$,
\begin{align}
    x^t_i \in \X = [-\xi, \, \xi]^d \label{eqn:confine_local_est}
\end{align}
where $\xi < \infty$. Thus,~\eqref{eqn:grad_exists} and~\eqref{eqn:confine_local_est} implies that, for all $i \in \H$ and $t \geq 0$,
\begin{align}
    \norm{\nabla Q_i(x^t_i)} \leq \max_{x \in \X} \norm{\nabla Q_i(x)} < \infty. \label{eqn:grad_less_inf}
\end{align}
Recall that $\nu_i \in \H$ in~\eqref{eqn:cge_nu_i}. Substituting from~\eqref{eqn:grad_less_inf} in~\eqref{eqn:cge_nu_i} we obtain that
\begin{align*}
    \norm{\cge_f \left\{ g^t_{i1}, \ldots, \, g^t_{in}\right\}} \leq (n-f) \max_{x \in \X} \norm{\nabla Q_{\nu_i}(x)}. 
\end{align*}
As $i$ is an arbitrary non-faulty agent, and $t$ is an arbitrary iteration, the above implies that $\norm{\cge_f \left\{ g^t_{i1}, \ldots, \, g^t_{in}\right\}} \leq \zeta$, where 
\[\zeta = (n-f) \max_{i \in \H} \max_{x \in \X} \norm{\nabla Q_{i}(x)},\]
for all $i \in \H$ and $t \geq 0$.
\end{proof}

\section{Appendix: Proof of Lemma~\ref{lem:growth}}
\label{app:lem_growth}

\noindent \fbox{\begin{minipage}{0.97\textwidth}
\begin{lemma*}[Restated]
Suppose that Assumptions~\ref{asp:finite},~\ref{asp:lipschitz} and~\ref{asp:step-size} hold true. Consider the algorithm presented in Section~\ref{sub:update}. There exist positive real values $\rho< 1$ and $\Gamma < \infty$ such that for each pair of non-faulty agent $i, \, j \in \H$, for all $k \in \{1, \ldots, \, d\}$ and $t \geq 1$,
\begin{align*}
    \mnorm{x^{t}_i[k] - x^{t}_j[k]} \leq \rho^t  \, \mnorm{\H} \, \max_{l \in \H}\mnorm{x^0_l[k]}  + \mnorm{\H} \Gamma ~ \sum_{r = 1}^{t} \eta_r \, \rho^{t - r} . 
\end{align*}
\end{lemma*}
\end{minipage}}
~

\begin{proof}
Our proof relies on the result in~\cite[Lemma~2]{su2020byzantine}. \\

Consider an arbitrary non-faulty agent $i \in \H$ and iteration $t \geq 0$.
Recall, from the update rule~\eqref{eqn:update} of the algorithm, that 
\begin{align*}
    x^{t+1}_i = P_{\X}\left(z^t_i - \eta_t \, \cge_f \left\{ g^t_{i1}, \ldots, \, g^t_{in}\right\}\right).
\end{align*}
Note that, by definition of $P_{\X}(\cdot)$ in~\eqref{eqn:proj_X} and~\eqref{eqn:alt_proj_X}, for all $k \in \{1, \ldots, \, d\}$,
\begin{align}
    x^{t+1}_i[k] = P_{[-\xi, \, \xi]}\left(z^t_i[k] - \eta_t \, \cge_f \left\{ g^t_{i1}, \ldots, \, g^t_{in}\right\} [k] \right). \label{eqn:update_k}
\end{align}
Upon substituting, for all $i$,
\[h^t_i = \cge_f \left\{ g^t_{i1}, \ldots, \, g^t_{in}\right\}\]
in~\eqref{eqn:update_k} we obtain that, for all $k \in \{1, \ldots, \, d\}$,
\begin{align}
    x^{t+1}_i[k] = P_{[-\xi, \, \xi]}\left(z^t_i[k] - \eta_t \, h^t_i[k]\right). \label{eqn:update_k_2}
\end{align}
Alternately, for all $k$,
\begin{align}
    x^{t+1}_i[k] = z^t_i[k] - \eta_t \, h^t_i[k] + P_{[-\xi, \, \xi]}\left(z^t_i[k] - \eta_t \, h^t_i[k]\right) - \left(z^t_i[k] - \eta_t \, h^t_i[k] \right). \label{eqn:update_k_3}
\end{align}
For all $i \in \H$ and $t$, we define the projection error to be
\begin{align}
    e^t_i = P_{\X}\left(z^t_i - \eta_t \, h^t_i\right) - \left(z^t_i - \eta_t \, h^t_i \right). \label{eqn:proj_error_i_0}
\end{align}
Therefore, for all $k \in \{1, \ldots, \, d\}$, 
\begin{align}
    e^t_i[k] = P_{[-\xi, ~ \xi]}\left(z^t_i[k] - \eta_t \, h^t_i[k]\right) - \left(z^t_i[k] - \eta_t \, h^t_i[k] \right). \label{eqn:proj_error_i}
\end{align}
Substituting from~\eqref{eqn:proj_error_i} in~\eqref{eqn:update_k_3} we obtain that, for all $k$,
\begin{align}
    x^{t+1}_i[k] = z^t_i[k] - \eta_t h^t_i[k] + e^t_i[k]. \label{eqn:update_k_4}
\end{align}
The update law in~\eqref{eqn:update_k_4} is similar to~\cite[Eqn.~(9)]{su2020byzantine}. Also, recall from Lemma~\ref{lem:bounded} that under Assumptions~\ref{asp:finite} and~\ref{asp:lipschitz}, $\norm{h^t_i} \leq \zeta < \infty$ for all $t$. Therefore, $\mnorm{h^t_i[k]} \leq \norm{h^t_i} \leq \zeta$ for all $k$ and $t$. We show below that for all $k$ and $t$, $\mnorm{e^t_i[k]} \leq \eta_t \zeta$.
The rest of the proof then follows immediately from the proof of~\cite[Lemma~2]{su2020byzantine}. \\

\noindent\hfil\rule{0.75\textwidth}{.4pt}\hfil
~\\

Consider an arbitrary $k \in \{1, \ldots, \, d\}$. Note from~\eqref{eqn:update_k_4} that
\begin{align}
    e^t_i[k] = x^{t+1}_i[k] - \left(z^t_i[k] - \eta_t h^t_i[k]\right). \label{eqn:proj_error_i_k}
\end{align}
Note that $z^t_i[k]$ can be written as a convex combination of non-faulty agents' estimates $\{x^t_{l}[k, ~ l \in \H]\}$ (see~\cite[Section VI.A]{su2020byzantine}). Specifically, as shown in~\eqref{eqn:cvx_z_1} of Section~\ref{sec:proof}), for all $t$,
\begin{align*}
    z^t_i[k] = \sum_{l \in \H}  \, \beta^t_{j,l}[k] \, x^t_{l}[k]. 
\end{align*}
where $\beta^t_{j, l}[k] \geq 0$ for all $l \in \H$ and $\sum_{l \in \H}  \, \beta^t_{j,l}[k] = 1$. Recall from~\eqref{eqn:update_k_2} that $x^t_l[k] \in [-\xi, \, \xi]$ for all $l \in \H$ and $t$. Therefore, the above implies that 
\begin{align}
    z^t_i[k] \in [-\xi, \, \xi], \quad \forall t, \, k. \label{eqn:lm_pf_zt_xi}
\end{align}
By the non-expansion property of projection onto a convex compact set~\cite[Lemma 1]{nedic2010constrained}, such as the set $\X \subset \R^d$, for all $v \in \R^d$ and $u \in \X$, 
\begin{align}
    \norm{P_{\X}(v) - u}^2 \leq \norm{v - u}^2 - \norm{P_{\X}(v) - v}^2. \label{eqn:proj_prop_pf}
\end{align}
The above property holds true even when $d = 1$. Now, recall from~\eqref{eqn:lm_pf_zt_xi} that $x^{t+1}_i[k] = P_{[-\xi, \, \xi]}\left(z^t_i[k] - \eta_t \, h^t_i[k]\right)$. As $z^t_i[k] \in [-\xi, \, \xi]$ (see~\eqref{eqn:lm_pf_zt_xi}), from~\eqref{eqn:proj_prop_pf} we obtain that 
\begin{align*}
    \mnorm{x^{t+1}_i[k] - z^t_i[k]}^2 \leq \mnorm{z^t_i[k] - \eta_t \, h^t_i[k] - z^t_i[k]}^2 - \mnorm{x^{t+1}_i[k] -  \left(z^t_i[k] - \eta_t \, h^t_i[k]\right)}^2
\end{align*}
Substituting from~\eqref{eqn:proj_error_i_k} in the above inequality implies that
\begin{align}
    \mnorm{x^{t+1}_i[k] - z^t_i[k] }^2 \leq \mnorm{\eta_t \, h^t_i[k]}^2 - \mnorm{e^t_{i}[k]}^2. \label{eqn:lm_pf_et_1}
\end{align}
From~\eqref{eqn:lm_pf_et_1} we obtain that
\begin{align*}
    \mnorm{e^t_{i}[k]}^2 \leq \mnorm{\eta_t \, h^t_i[k]}^2 - \mnorm{x^{t+1}_i[k] - z^t_i[k] }^2 \leq \mnorm{\eta_t \, h^t_i[k]}^2.
\end{align*}
Therefore, 
\begin{align}
    \mnorm{e^t_{i}[k]} \leq \eta_t \,\mnorm{ h^t_i[k]}. \label{eqn:lem_pf_et_bnd}
\end{align}
Recall that under Assumptions~\ref{asp:finite} and~\ref{asp:lipschitz}, owing to Lemma~\ref{lem:bounded}, $\norm{h^t_i} \leq \zeta < \infty$ for all $t$. Therefore, $\mnorm{h^t_i[k]} \leq \norm{h^t_i} \leq \zeta$ for all $k$ and $t$. Substituting this in~\eqref{eqn:lem_pf_et_bnd} implies that
\begin{align*}
    \mnorm{e^t_{i}[k]} \leq \eta_t \,\zeta. 
\end{align*}
The above holds true for all $k \in \{1, \ldots, \, d\}$ and $t \geq 0$. 
\end{proof}

%% file: app_proof_claims.tex
\section{Appendix: Proof of Claim~\ref{clm:bnd_inf_sum}}
\label{app:clm_bnd}

\noindent \fbox{\begin{minipage}{0.97\textwidth}
\begin{claim*}[Restated]
If Assumptions~\ref{asp:finite},~\ref{asp:lipschitz}  and~\ref{asp:step-size} hold true then for all $k$,
\begin{align*}
    \sum_{t = 0}^{\infty} \eta_t \, \max_{i, \, j \in \H} \mnorm{x^t_i[k] - x^t_j[k]} < \infty. 
\end{align*}
\end{claim*}
\end{minipage}}
~\\

\begin{proof}
Recall from Lemma~\ref{lem:growth} that, under Assumptions~\ref{asp:lipschitz} and~\ref{asp:step-size}, there exists $\rho \in (0, \, 1)$ and $\Gamma < \infty$ such that for each pair of non-faulty agent $i, \, j \in \H$, for all $k \in \{1, \ldots, \, d\}$ and $t \geq 0$,
\begin{align*}
    \mnorm{x^{t}_i[k] - x^{t}_j[k]} \leq \rho^t  \, \mnorm{\H} \, \max_{l \in \H}\mnorm{x^0_l[k]}  + \mnorm{\H} \Gamma ~ \sum_{r = 0}^{t} \eta_r \, \rho^{t - r} . 
\end{align*}
Therefore, for all $t \geq 0$ and $k \in \{1, \ldots, \, d\}$, 
\begin{align}
    \max_{i, \, j \in \H} \mnorm{x^t_i[k] - x^t_j[k]} \leq  \rho^t  \, \mnorm{\H} \, \max_{l \in \H}\mnorm{x^0_l[k]}  + \mnorm{\H} \Gamma ~ \sum_{r = 0}^{t} \eta_r \, \rho^{t - r}. \label{eqn:max_growth_0}
\end{align}
Upon multiplying both sides in~\eqref{eqn:max_growth_0} by $\eta_t$, and then taking summation over $t = 0, \, 1, \ldots $ we obtain that, for all $k \in \{1, \ldots, \, d\}$,
\begin{align}
    \sum_{t = 0}^{\infty} \eta_t \max_{i, \, j \in \H} \mnorm{x^t_i[k] - x^t_j[k]} \leq  \mnorm{\H} \, \max_{l \in \H}\mnorm{x^0_l[k]} \, \sum_{t = 0}^{\infty} \eta_t \rho^t  + \mnorm{\H} \Gamma \, \sum_{t = 0}^{\infty} \eta_t \sum_{r = 0}^{t} \eta_r \, \rho^{t - r}. \label{eqn:max_growth}
\end{align}
Recall that, under Assumption~\ref{asp:step-size}, $\eta_t \leq \eta_0 < \infty$ for all $t \geq 0$, and that $0 \leq \rho < 1$. Therefore, 
\begin{align}
    \sum_{t = 0}^{\infty} \eta_t \, \rho^t  \leq \eta_0 \, \sum_{t = 0}^{\infty} \rho^t = \frac{\eta_0}{1 - \rho} < \infty \label{eqn:eta_rho_sum_1}.
\end{align}
As $x^0_i$ is a finite point for all $i \in \H$,~\eqref{eqn:eta_rho_sum_1} implies that the first term in~\eqref{eqn:max_growth}, i.e., 
\begin{align}
    \mnorm{\H} \, \max_{l \in \H}\mnorm{x^0_l[k]} \, \sum_{t = 0}^{\infty} \eta_t \rho^t < \infty. \label{eqn:sub_max_growth_1}
\end{align}
We show below that the summation in the second term of~\eqref{eqn:max_growth} is also finite, i.e., $\sum_{t = 0}^{\infty} \eta_t \sum_{r = 0}^{t} \eta_r \, \rho^{t - r} < \infty$.. \\

For a real value $a$, we let $\floor{a}$ denote the largest integer smaller than or equal to $a$. For an arbitrary $t \geq 2$, we can write
\begin{align}
    \sum_{r = 0}^{t} \eta_r \, \rho^{t - r} = \sum_{r = 0}^{\floor{t/2} } \eta_r \, \rho^{t - r} + \sum_{r = \floor{t/2} + 1}^{t} \eta_r \, \rho^{t - r}. \label{eqn:sigma_1}
\end{align}
Note that, as $\eta_{t+1} \leq \eta_{t}$ for all $t \geq 0$ (cf.~Assumption~\ref{asp:step-size}) and $\rho \in (0,\, 1)$,
\begin{align}
    \sum_{r = 0}^{\floor{t/2} } \eta_r \, \rho^{t - r} \leq \eta_0 \, \sum_{r = 0}^{\floor{t/2}} \rho^{t - r} \leq \eta_0 \, \rho^{t/2} \, \sum_{r = 0}^{\infty}\rho^r = \frac{\eta_0 \, \rho^{t/2}}{1 - \rho}. \label{eqn:sub_sigma_1}
\end{align}
Similarly,
\begin{align}
    \sum_{r = \floor{t/2} + 1}^{t} \eta_r \, \rho^{t - r} \leq \eta_{(\floor{t/2} + 1)} \, \sum_{r = 0}^{\infty}\rho^r = \frac{\eta_{(\floor{t/2} + 1)}}{1 - \rho}. \label{eqn:sub_sigma_2}
\end{align}
Upon substituting from~\eqref{eqn:sub_sigma_1} and~\eqref{eqn:sub_sigma_2} in~\eqref{eqn:sigma_1} we obtain that, for all $t \geq 2$,
\begin{align}
    \sum_{r = 0}^{t} \eta_r \, \rho^{t - r} \leq  \left( \eta_0 \, \rho^{t/2}   + \eta_{(\floor{t/2} + 1)}  \right)\left(\frac{1}{1 - \rho}\right). \label{eqn:sigma_2}
\end{align}
From~\eqref{eqn:sigma_2} we obtain that
\begin{align}
    \sum_{t = 2}^{\infty} \eta_t \sum_{r = 0}^{t} \eta_r \, \rho^{t - r} \leq \left(\frac{1}{1 - \rho}\right)\left( \eta_0 \, \sum_{t = 2}^\infty \eta_t \rho^{t/2} +  \sum_{t = 2}^{\infty} \eta_t \, \eta_{(\floor{t/2} + 1)} \right).\label{eqn:sigma_3}
\end{align}
Note that, as $\rho \in (0, \, 1)$ and $\eta_{t+1} \leq \eta_t$ for all $t$ (see Assumption~\ref{asp:step-size}),
\begin{align}
    \sum_{t = 2}^\infty \eta_t \rho^{t/2} \leq \eta_2  \sum_{t = 2}^\infty \left(\sqrt{\rho}\right)^t \leq \eta_2  \sum_{t = 0}^\infty \left(\sqrt{\rho}\right)^t \leq \frac{\eta_2}{1 - \sqrt{\rho}} < \infty. \label{eqn:sigma_3_1} 
\end{align}
Also, by Assumption~\ref{asp:step-size}, as $\eta_{t+1} \leq \eta_t$ for all $t$ and $\sum_{t = 0}^\infty \eta^2_t < \infty$,
\begin{align}
    \sum_{t = 2}^{\infty} \eta_t \, \eta_{(\floor{t/2} + 1)} \leq \sum_{t = 2}^\infty \eta^2_{\floor{t/2}} = 2 \sum_{t = 1}^\infty \eta^2_t < \infty. \label{eqn:sigma_3_2} 
\end{align}
Substituting from~\eqref{eqn:sigma_3_1} and~\eqref{eqn:sigma_3_2} in~\eqref{eqn:sigma_3} implies that
\begin{align*}
    \sum_{t = 2}^{\infty} \eta_t \sum_{r = 0}^{t} \eta_r \, \rho^{t - r} < \infty. 
\end{align*}
As $\eta_t < \infty$ for all $t$, the above implies that
\begin{align}
    \mnorm{\H} \Gamma \, \sum_{t = 0}^{\infty} \eta_t \sum_{r = 0}^{t} \eta_r \, \rho^{t - r} = \mnorm{\H} \Gamma \, \left( \eta^2_0 + \rho \eta_0 \eta_1 + \eta^2_1 + \sum_{t = 2}^{\infty} \eta_t \sum_{r = 0}^{t} \eta_r \, \rho^{t - r} \right) < \infty. \label{eqn:sigma_4}
\end{align}
~

Finally, substituting from~\eqref{eqn:sub_max_growth_1} and~\eqref{eqn:sigma_4} in~\eqref{eqn:max_growth} we obtain that, for all $k$,
\begin{align}
    \sum_{t = 0}^{\infty} \eta_t \max_{i, \, j \in \H} \mnorm{x^t_i[k] - x^t_j[k]} < \infty. \label{eqn:max_growth_2}
\end{align}
Hence, the proof.
\end{proof}

\section{Appendix: Proof of Claim~\ref{clm:ht_ij}}
\label{app:clm_ht}

We present below a proof of Claim~\ref{clm:ht_ij}, which is a part of Theorem~\ref{thm:main}'s proof presented in Section~\ref{sec:proof}. For convenience, we re-state the claim below. \\

\noindent \fbox{\begin{minipage}{0.97\textwidth}
\begin{claim*}[Restated]
If the $2f$-redundancy property and Assumption~\ref{asp:lipschitz} hold true then for two arbitrary non-faulty agents $i, \, j \in \H$, for all $k \in \{1, \ldots, \, d\}$,
\begin{align*}
    \left(x^t_i[k] - x^*[k]\right) h^t_j[k] & \geq \sum_{l \in \H} \left(x^t_i[k] - x^*[k]\right) \nabla Q_l(x^t_i)[k] - 2 \mu f \mnorm{x^t_i[k] - x^*[k]} \norm{x^t_i - x^*} \nonumber \\
    & - \mu \left( n + 2f \right) \, \mnorm{x^t_i[k] - x^*[k]}  \max_{l \in \H} \norm{x^t_l - x^t_i}. 
\end{align*}
\end{claim*}
\end{minipage}}
~

\begin{proof}
Consider an arbitrary non-faulty agent $j \in \H$. Recall that 
\[h^t_j = \cge_f \left\{ g^t_{j1}, \ldots, \, g^t_{jn}\right\}.\]
From the definition of $\cge_f$ in~\eqref{eqn:cgc} note that, as there are at most $f$ Byzantine faulty agents, for each iteration $t$ there exist disjoint sets $\H^t_j  \subseteq \H$ and $\B^t_j \subseteq \B$ with $\mnorm{\H^t_j} \geq \mnorm{\H} - f$, $\mnorm{\B^t_j} \leq f$, and $\mnorm{\H^t_i \cup \B^t_j} = n-f$, such that 
\begin{align*}
    h^t_j = \sum_{l \in \H^t_j} g^t_{jl} + \sum_{l \in \B^t_j} g^t_{jl}.
\end{align*}
Recall that $g^t_{jl} = \nabla Q_l(x^t_l)$ for all $l \in \H$. Substituting this above we obtain that
\begin{align*}
    h^t_j  = \sum_{l \in \H^t_j} \nabla Q_l(x^t_l) + \sum_{l \in \B^t_j} g^t_{jl} = \sum_{l \in \H} \nabla Q_l(x^t_l) - \sum_{l \in \H \setminus \H^t_j}\nabla Q_l(x^t_l) +  \sum_{l \in \B^t_j} g^t_{jl}.
\end{align*}
Therefore, for each $k \in \{1, \ldots, \, d\}$,
\begin{align*}
    h^t_j[k]  = \sum_{l \in \H} \nabla Q_l(x^t_l)[k] - \sum_{l \in \H \setminus \H^t_j}\nabla Q_l(x^t_l)[k] +  \sum_{l \in \B^t_j} g^t_{jl}[k]. 
\end{align*}
Now, consider another arbitrary non-faulty agent $i \in \H$. For each $k$, upon multiplying both sides above by $\left(x^t_i[k] - x^*[k]\right)$ we obtain that
\begin{align}
    \left(x^t_i[k] - x^*[k]\right) h^t_j[k]  = & \sum_{l \in \H} \left(x^t_i[k] - x^*[k]\right) \nabla Q_l(x^t_l)[k] \label{eqn:ht_j_0} \\
    & - \sum_{l \in \H \setminus \H^t_j} \left(x^t_i[k] - x^*[k]\right) \nabla Q_l(x^t_l)[k] +  \sum_{l \in \B^t_j} \left(x^t_i[k] - x^*[k]\right) g^t_{jl}[k].  \nonumber
\end{align}
Note that 
\[ \left(x^t_i[k] - x^*[k]\right) \nabla Q_l(x^t_l)[k] \leq \mnorm{x^t_i[k] - x^*[k]} \, \mnorm{Q_l(x^t_l)[k]}, ~ \forall l \in \H \setminus \H^t_j \]
and similarly, 
\[\left(x^t_i[k] - x^*[k]\right) g^t_{jl}[k] \geq - \mnorm{x^t_i[k] - x^*[k]} \, \mnorm{g^t_{jl}[k]}, ~ \forall l \in \B^t_j.\]
Substituting from above in~\eqref{eqn:ht_j_0} we obtain that
\begin{align}
    \left(x^t_i[k] - x^*[k]\right) h^t_j[k] \geq & \sum_{l \in \H} \left(x^t_i[k] - x^*[k]\right) \nabla Q_l(x^t_l)[k]  \label{eqn:ht_j_1} \\
    & - \mnorm{x^t_i[k] - x^*[k]} \left(\sum_{l \in \H \setminus \H^t_j} \mnorm{\nabla Q_l(x^t_l)[k]} +  \sum_{l \in \B^t_j} \mnorm{g^t_{jl}[k]} \right). \nonumber
\end{align}
Note that for a vector $v \in \R^d$ the absolute value of its each element is less than or equal to its norm, i.e., for each $k$, $\mnorm{v[k]} \leq \norm{v}$.
Therefore, 
\[ \mnorm{\nabla Q_l(x^t_l)[k]} \leq \norm{\nabla Q_l(x^t_l)}, \quad \forall l \in \B^t_j, \]
and
\[\mnorm{g^t_{jl}[k]} \leq \norm{g^t_{jl}}, \quad \forall l \in \B^t_j.\]
Substituting from above in~\eqref{eqn:ht_j_1} we obtain that
\begin{align}
    \left(x^t_i[k] - x^*[k]\right) h^t_j[k] \geq & \sum_{l \in \H} \left(x^t_i[k] - x^*[k]\right) \nabla Q_l(x^t_l)[k] \nonumber \\
    & - \mnorm{x^t_i[k] - x^*[k]} \left(\sum_{l \in \H \setminus \H^t_j} \norm{\nabla Q_l(x^t_l)} +  \sum_{l \in \B^t_j} \norm{g^t_{jl}} \right). \label{eqn:ht_j_2}
\end{align}
Note from the definition of $\cge_f$ in~\eqref{eqn:cgc} that for each $l \in \B^t_j$ there exists a unique $l \in \H \setminus \H^t_j$ such that $\norm{g^t_{jl}} \leq \norm{\nabla Q_l(x^t_l)}$. Therefore, 
\begin{align*}
    \sum_{l \in \B^t_j} \norm{g^t_{jl}} \leq \sum_{l \in \H \setminus \H^t_j} \norm{\nabla Q_l(x^t_l)}.
\end{align*}
Substituting from above in~\eqref{eqn:ht_j_2} we obtain that
\begin{align}
    \left(x^t_i[k] - x^*[k]\right) h^t_j[k]  \geq & \sum_{l \in \H} \left(x^t_i[k] - x^*[k]\right) \nabla Q_l(x^t_l)[k] \nonumber \\
    & - 2 \mnorm{x^t_i[k] - x^*[k]} \sum_{l \in \H \setminus \H^t_j} \norm{\nabla Q_l(x^t_l)}. \label{eqn:ht_j_3}
\end{align}
For all $l \in \H$, the first term on the right-hand side in~\eqref{eqn:ht_j_3} above can be written as
\begin{align}
    \left(x^t_i[k] - x^*[k]\right) \nabla Q_l(x^t_l)[k] = & \left(x^t_i[k] - x^*[k]\right) \nabla Q_l(x^t_i)[k] \nonumber \\ 
    & + \left(x^t_i[k] - x^*[k]\right) \left( \nabla Q_l(x^t_l)[k] -  \nabla Q_l(x^t_i)[k] \right). \label{eqn:above_ht_j_3}
\end{align}
Upon substituting 
\[\left(x^t_i[k] - x^*[k]\right) \left( \nabla Q_l(x^t_l)[k] - \nabla Q_l(x^t_i)[k] \right) \geq - \mnorm{x^t_i[k] - x^*[k]} \norm{\nabla Q_l(x^t_l) - \nabla Q_l(x^t_i)}\]
in~\eqref{eqn:above_ht_j_3},
we obtain that
\begin{align*}
    \left(x^t_i[k] - x^*[k]\right) \nabla Q_l(x^t_l)[k] \geq & \left(x^t_i[k] - x^*[k]\right) \nabla Q_l(x^t_i)[k] \\ 
    & - \mnorm{x^t_i[k] - x^*[k]} \norm{\nabla Q_l(x^t_l) - \nabla Q_l(x^t_i)}.
\end{align*}
Substituting from above in~\eqref{eqn:ht_j_3} we obtain that
\begin{align}
    & \left(x^t_i[k] - x^*[k]\right) h^t_j[k]  \geq \sum_{l \in \H} \left(x^t_i[k] - x^*[k]\right) \nabla Q_l(x^t_i)[k]  \nonumber \\
    & - \mnorm{x^t_i[k] - x^*[k]} \sum_{l \in \H} \norm{\nabla Q_l(x^t_l) - \nabla Q_l(x^t_i)} - 2  \sum_{l \in \H \setminus \H^t_j} \mnorm{x^t_i[k] - x^*[k]} \norm{\nabla Q_l(x^t_l)}. \label{eqn:ht_j_5}
\end{align}
Using the triangle inequality, for each $l \in \H \setminus \H^t_j$, in the last term on the right-hand side in~\eqref{eqn:ht_j_5} above we obtain that
\begin{align*}
    \norm{\nabla Q_l(x^t_l)} \leq \norm{\nabla Q_l(x^t_i)} + \norm{\nabla Q_l(x^t_l) - \nabla Q_l(x^t_i)}.
\end{align*}
Substituting from above in~\eqref{eqn:ht_j_5} we obtain that
\begin{align}
    & \left(x^t_i[k] - x^*[k]\right) h^t_j[k]  \geq \sum_{l \in \H} \left(x^t_i[k] - x^*[k]\right) \nabla Q_l(x^t_i)[k] - 2 \sum_{l \in \H \setminus \H^t_j} \mnorm{x^t_i[k] - x^*[k]} \norm{\nabla Q_l(x^t_i)} \nonumber \\
    & - \mnorm{x^t_i[k] - x^*[k]} \sum_{l \in \H} \norm{ \nabla Q_l(x^t_l) - \nabla Q_l(x^t_i)} - 2  \sum_{l \in \H \setminus \H^t_j} \mnorm{x^t_i[k] - x^*[k]} \norm{\nabla Q_l(x^t_l) - \nabla Q_l(x^t_i)}. \label{eqn:ht_j_6}
\end{align}
From Lipschitz continuity of all the non-faulty gradients, i.e., Assumption~\ref{asp:lipschitz}, we obtain that, for all $l \in \H$,
\[\norm{\nabla Q_l(x^t_l) - \nabla Q_l(x^t_i)} \leq \mu \norm{x^t_l - x^t_i} \leq \mu \max_{l \in \H} \norm{x^t_l - x^t_i}.\]
Substituting from above in~\eqref{eqn:ht_j_6} we obtain that
\begin{align*}
    & \left(x^t_i[k] - x^*[k]\right) h^t_j[k]  \geq \sum_{l \in \H} \left(x^t_i[k] - x^*[k]\right) \nabla Q_l(x^t_i)[k] - 2 \sum_{l \in \H \setminus \H^t_j} \mnorm{x^t_i[k] - x^*[k]} \norm{\nabla Q_l(x^t_i)} \nonumber \\
    & - \mu \left( \mnorm{\H} + 2 \mnorm{\H \setminus \H^t_j} \right) \, \mnorm{x^t_i[k] - x^*[k]}  \max_{l \in \H} \norm{x^t_l - x^t_i}. 
\end{align*}
Recall that $\mnorm{\H \setminus \H^t_j} \leq f$ and $\mnorm{\H} \leq n$. Substituting this above we obtain that
\begin{align}
     \left(x^t_i[k] - x^*[k]\right) h^t_j[k] \geq & \sum_{l \in \H} \left(x^t_i[k] - x^*[k]\right) \nabla Q_l(x^t_i)[k] - 2 \sum_{l \in \H \setminus \H^t_j} \mnorm{x^t_i[k] - x^*[k]} \norm{\nabla Q_l(x^t_i)} \nonumber \\
    & - \mu \left( n + 2f \right) \, \mnorm{x^t_i[k] - x^*[k]}  \max_{l \in \H} \norm{x^t_l - x^t_i}. \label{eqn:ht_j_7}
\end{align}
We know that under the $2f$-redundancy property~\cite{gupta2020fault_podc}, $\nabla Q_l(x^*) = 0$ for all $l \in \H$. Thus, the Lipschitz continuity assumption (i.e., Assumption~\ref{asp:lipschitz}) implies that
\[\norm{\nabla Q_l(x^t_i)} \leq \mu \norm{x^t_i - x^*}, ~ \forall l \in \H.\]
Upon substituting the above in~\eqref{eqn:ht_j_7} we obtain that
\begin{align*}
     \left(x^t_i[k] - x^*[k]\right) h^t_j[k] \geq & \sum_{l \in \H} \left(x^t_i[k] - x^*[k]\right) \nabla Q_l(x^t_i)[k] - 2 \mu \mnorm{\H \setminus \H^t_j} \mnorm{x^t_i[k] - x^*[k]} \norm{x^t_i - x^*} \nonumber \\
    & - \mu \left( n + 2f \right) \, \mnorm{x^t_i[k] - x^*[k]}  \max_{l \in \H} \norm{x^t_l - x^t_i}. 
\end{align*}
As $\mnorm{\H \setminus \H^t_j} \leq f$, from above we obtain that
\begin{align}
     \left(x^t_i[k] - x^*[k]\right) h^t_j[k] \geq & \sum_{l \in \H} \left(x^t_i[k] - x^*[k]\right) \nabla Q_l(x^t_i)[k] - 2 \mu f \mnorm{x^t_i[k] - x^*[k]} \norm{x^t_i - x^*} \nonumber \\
    & - \mu \left( n + 2f \right) \, \mnorm{x^t_i[k] - x^*[k]}  \max_{l \in \H} \norm{x^t_l - x^t_i}.  \label{eqn:clm_l_sum}
\end{align}
Note that in the right hand side above, 
\[ \sum_{l \in \H} \left(x^t_i[k] - x^*[k]\right) \nabla Q_l(x^t_i)[k] =  \left(x^t_i[k] - x^*[k]\right) \sum_{l \in \H} \nabla Q_l(x^t_i)[k].\]
Recall, from~\eqref{eqn:def_cost_h}, that $Q_\H(x) = (1/\mnorm{\H}) \sum_{i \in H} Q_i(x)$ for all $x \in \R^d$. Thus, $$\sum_{l \in \H} \nabla Q_l(x^t_i)[k] = \mnorm{\H} \nabla Q_{\H}(x^t_i)[k],$$
and
\[ \sum_{l \in \H} \left(x^t_i[k] - x^*[k]\right) \nabla Q_l(x^t_i)[k] =  \mnorm{\H} \left(x^t_i[k] - x^*[k]\right) \nabla Q_\H(x^t_i)[k].\]
Substituting from above in~\eqref{eqn:clm_l_sum} we obtain that
\begin{align*}
    \left(x^t_i[k] - x^*[k]\right) h^t_j[k] & \geq  \mnorm{\H} \left(x^t_i[k] - x^*[k]\right) \nabla Q_\H(x^t_i)[k] - 2 \mu f \mnorm{x^t_i[k] - x^*[k]} \norm{x^t_i - x^*} \\
    & - \mu \left( n + 2f \right) \, \mnorm{x^t_i[k] - x^*[k]}  \max_{l \in \H} \norm{x^t_l - x^t_i}. 
\end{align*}
Hence, the proof.
\end{proof}

\section{Appendix: Proof of Claim~\ref{clm:zt_ij}}
\label{app:clm_zt}

We present below a proof of Claim~\ref{clm:zt_ij}, which is a part of Theorem~\ref{thm:main}'s proof presented in Section~\ref{sec:proof}. For convenience, we re-state the claim below.  \\

\noindent \fbox{\begin{minipage}{0.97\textwidth}
\begin{claim*}
If Assumptions~\ref{asp:finite} and~\ref{asp:lipschitz} hold true then for two arbitrary non-faulty agents $i, \, j \in \H$, for all $k \in \{1, \ldots, \, d\}$, 
\begin{align*}
    \mnorm{\left(z^t_{j}[k] - x^t_i[k]  \right)  h^t_j[k]} \leq \zeta \, \max_{l \in \H} \mnorm{x^t_l[k] - x^t_i[k]}
\end{align*}
where recall from Lemma~\ref{lem:bounded} that $\zeta \in (0, \, \infty)$ such that $\norm{h^t_j} \leq \zeta, ~  \forall j \in \H$.
\end{claim*}
\end{minipage}}
~

\begin{proof}
Consider an arbitrary $k \in \{1, \ldots, \, d\}$. Recall from~\eqref{eqn:cvx_z_1} that 
\begin{align*}
    z^t_j[k] = \sum_{l \in \H}  \, \beta^t_{j,l}[k] \, x^t_{l}[k]. 
\end{align*}
where $\beta^t_{j, l}[k] \geq 0$ for all $l \in \H$ and $\sum_{l \in \H}  \, \beta^t_{j,l}[k] = 1$. Upon subtracting $x^t_i[k]$ on both sides above, we obtain that
\begin{align*}
    z^t_j[k] -  x^t_i[k] = \sum_{l \in \H}  \, \beta^t_{j,l}[k] \, (x^t_{l}[k] - x^t_i[k]). 
\end{align*}
From triangle inequality, and the fact that $\beta^t_{j, l}[k] \geq 0$ for all $l \in \H$, we obtain that
\begin{align}
    \mnorm{z^t_j[k] -  x^t_i[k]} \leq \sum_{l \in \H}  \, \beta^t_{j,l}[k] \, \mnorm{x^t_{l}[k] - x^t_i[k]} \leq \max_{l \in \H} \mnorm{x^t_l[k] - x^t_i[k]}. \label{eqn:clm_ct_pf_1}
\end{align}
Under Assumptions~\ref{asp:finite} and~\ref{asp:lipschitz}, owing to Lemma~\ref{lem:bounded},
\[\norm{h^t_j} \leq \zeta, \quad \forall j \in \H, \, t.\]
Therefore, 
\begin{align}
    \mnorm{h^t_j[k]} \leq \norm{h^t_j} \leq \zeta. \label{eqn:clm_ct_pf_2}
\end{align}
Finally, note that
\begin{align*}
    \mnorm{\left(z^t_{j}[k] - x^t_i[k]  \right)  h^t_j[k]} \leq \mnorm{z^t_j[k] -  x^t_i[k]} \, \mnorm{h^t_j[k]}.
\end{align*}
Substituting from~\eqref{eqn:clm_ct_pf_1} and~\eqref{eqn:clm_ct_pf_2} in the inequality above proves the claim.
\end{proof}

%% file: main.bbl
\begin{thebibliography}{10}

\bibitem{alistarh2018byzantine}
Dan Alistarh, Zeyuan Allen-Zhu, and Jerry Li.
\newblock Byzantine stochastic gradient descent.
\newblock In {\em Advances in Neural Information Processing Systems}, pages
  4618--4628, 2018.

\bibitem{anthonisse1977exponential}
Jac~M Anthonisse and Henk Tijms.
\newblock Exponential convergence of products of stochastic matrices.
\newblock {\em Journal of Mathematical Analysis and Applications},
  59(2):360--364, 1977.

\bibitem{bernstein2018signsgd}
Jeremy Bernstein, Jiawei Zhao, Kamyar Azizzadenesheli, and Anima Anandkumar.
\newblock signsgd with majority vote is communication efficient and {Byzantine}
  fault tolerant.
\newblock {\em arXiv preprint arXiv:1810.05291}, 2018.

\bibitem{bertsekas1989parallel}
Dimitri~P Bertsekas and John~N Tsitsiklis.
\newblock {\em Parallel and distributed computation: numerical methods},
  volume~23.
\newblock Prentice hall Englewood Cliffs, NJ, 1989.

\bibitem{blanchard2017machine}
Peva Blanchard, Rachid Guerraoui, Julien Stainer, et~al.
\newblock Machine learning with adversaries: {Byzantine} tolerant gradient
  descent.
\newblock In {\em Advances in Neural Information Processing Systems}, pages
  119--129, 2017.

\bibitem{bottou1998online}
L{\'e}on Bottou.
\newblock Online learning and stochastic approximations.
\newblock {\em On-line learning in neural networks}, 17(9):142, 1998.

\bibitem{bottou2018optimization}
L{\'e}on Bottou, Frank~E Curtis, and Jorge Nocedal.
\newblock Optimization methods for large-scale machine learning.
\newblock {\em Siam Review}, 60(2):223--311, 2018.

\bibitem{boyd2011distributed}
Stephen Boyd, Neal Parikh, Eric Chu, Borja Peleato, Jonathan Eckstein, et~al.
\newblock Distributed optimization and statistical learning via the alternating
  direction method of multipliers.
\newblock {\em Foundations and Trends{\textregistered} in Machine learning},
  3(1):1--122, 2011.

\bibitem{boyd2004convex}
Stephen Boyd and Lieven Vandenberghe.
\newblock {\em Convex optimization}.
\newblock Cambridge university press, 2004.

\bibitem{cao2019distributed}
Xinyang Cao and Lifeng Lai.
\newblock Distributed gradient descent algorithm robust to an arbitrary number
  of byzantine attackers.
\newblock {\em IEEE Transactions on Signal Processing}, 67(22):5850--5864,
  2019.

\bibitem{charikar2017learning}
Moses Charikar, Jacob Steinhardt, and Gregory Valiant.
\newblock Learning from untrusted data.
\newblock In {\em Proceedings of the 49th Annual ACM SIGACT Symposium on Theory
  of Computing}, pages 47--60, 2017.

\bibitem{chen2018resilient}
Yuan Chen, Soummya Kar, and Jose~MF Moura.
\newblock Resilient distributed estimation through adversary detection.
\newblock {\em IEEE Transactions on Signal Processing}, 66(9):2455--2469, 2018.

\bibitem{chen2017distributed}
Yudong Chen, Lili Su, and Jiaming Xu.
\newblock Distributed statistical machine learning in adversarial settings:
  {Byzantine} gradient descent.
\newblock {\em Proceedings of the ACM on Measurement and Analysis of Computing
  Systems}, 1(2):44, 2017.

\bibitem{chong2015observability}
Michelle~S Chong, Masashi Wakaiki, and Joao~P Hespanha.
\newblock Observability of linear systems under adversarial attacks.
\newblock In {\em American Control Conference}, pages 2439--2444. IEEE, 2015.

\bibitem{duchi2011dual}
John~C Duchi, Alekh Agarwal, and Martin~J Wainwright.
\newblock Dual averaging for distributed optimization: Convergence analysis and
  network scaling.
\newblock {\em IEEE Transactions on Automatic control}, 57(3):592--606, 2011.

\bibitem{gupta2020byzantine_thinh}
Nirupam Gupta, Thinh~T. Doan, and Nitin~H. Vaidya.
\newblock Byzantine fault-tolerance in decentralized optimization under minimal
  redundancy, 2020.

\bibitem{gupta2019byzantine}
Nirupam Gupta and Nitin~H Vaidya.
\newblock Byzantine fault tolerant distributed linear regression.
\newblock {\em arXiv preprint arXiv:1903.08752}, 2019.

\bibitem{gupta2020fault_podc}
Nirupam Gupta and Nitin~H Vaidya.
\newblock Fault-tolerance in distributed optimization: The case of redundancy.
\newblock In {\em Proceedings of the 39th Symposium on Principles of
  Distributed Computing}, pages 365--374, 2020.

\bibitem{gupta2020resilience}
Nirupam Gupta and Nitin~H Vaidya.
\newblock Resilience in collaborative optimization: Redundant and independent
  cost functions.
\newblock {\em arXiv preprint arXiv:2003.09675}, 2020.

\bibitem{kairouz2019advances}
Peter Kairouz, H~Brendan McMahan, Brendan Avent, Aur{\'e}lien Bellet, Mehdi
  Bennis, Arjun~Nitin Bhagoji, Keith Bonawitz, Zachary Charles, Graham Cormode,
  Rachel Cummings, et~al.
\newblock Advances and open problems in federated learning.
\newblock {\em arXiv preprint arXiv:1912.04977}, 2019.

\bibitem{lamport1982byzantine}
Leslie Lamport, Robert Shostak, and Marshall Pease.
\newblock The {Byzantine} generals problem.
\newblock {\em ACM Transactions on Programming Languages and Systems (TOPLAS)},
  4(3):382--401, 1982.

\bibitem{liu2021approximate}
Shuo Liu, Nirupam Gupta, and Nitin~H. Vaidya.
\newblock Approximate byzantine fault-tolerance in distributed optimization,
  2021.

\bibitem{nedic2009distributed}
Angelia Nedic and Asuman Ozdaglar.
\newblock Distributed subgradient methods for multi-agent optimization.
\newblock {\em IEEE Transactions on Automatic Control}, 54(1):48--61, 2009.

\bibitem{nedic2010constrained}
Angelia Nedic, Asuman Ozdaglar, and Pablo~A Parrilo.
\newblock Constrained consensus and optimization in multi-agent networks.
\newblock {\em IEEE Transactions on Automatic Control}, 55(4):922--938, 2010.

\bibitem{pajic2014robustness}
Miroslav Pajic, James Weimer, Nicola Bezzo, Paulo Tabuada, Oleg Sokolsky, Insup
  Lee, and George~J Pappas.
\newblock Robustness of attack-resilient state estimators.
\newblock In {\em ICCPS'14: ACM/IEEE 5th International Conference on
  Cyber-Physical Systems (with CPS Week 2014)}, pages 163--174. IEEE Computer
  Society, 2014.

\bibitem{rabbat2004distributed}
Michael Rabbat and Robert Nowak.
\newblock Distributed optimization in sensor networks.
\newblock In {\em Proceedings of the 3rd international symposium on Information
  processing in sensor networks}, pages 20--27, 2004.

\bibitem{raffard2004distributed}
Robin~L Raffard, Claire~J Tomlin, and Stephen~P Boyd.
\newblock Distributed optimization for cooperative agents: Application to
  formation flight.
\newblock In {\em 2004 43rd IEEE Conference on Decision and Control (CDC)(IEEE
  Cat. No. 04CH37601)}, volume~3, pages 2453--2459. IEEE, 2004.

\bibitem{rudin1964principles}
Walter Rudin.
\newblock {\em Principles of mathematical analysis}, volume~3.
\newblock McGraw-hill New York, 1964.

\bibitem{su2018finite}
Lili Su and Shahin Shahrampour.
\newblock Finite-time guarantees for {Byzantine}-resilient distributed state
  estimation with noisy measurements.
\newblock {\em arXiv preprint arXiv:1810.10086}, 2018.

\bibitem{su2016fault}
Lili Su and Nitin~H Vaidya.
\newblock Fault-tolerant multi-agent optimization: optimal iterative
  distributed algorithms.
\newblock In {\em Proceedings of the 2016 ACM symposium on principles of
  distributed computing}, pages 425--434. ACM, 2016.

\bibitem{su2016robust}
Lili Su and Nitin~H Vaidya.
\newblock Robust multi-agent optimization: coping with {Byzantine} agents with
  input redundancy.
\newblock In {\em International Symposium on Stabilization, Safety, and
  Security of Distributed Systems}, pages 368--382. Springer, 2016.

\bibitem{su2020byzantine}
Lili Su and Nitin~H Vaidya.
\newblock Byzantine-resilient multi-agent optimization.
\newblock {\em IEEE Transactions on Automatic Control}, 2020.

\bibitem{sundaram2018distributed}
Shreyas Sundaram and Bahman Gharesifard.
\newblock Distributed optimization under adversarial nodes.
\newblock {\em IEEE Transactions on Automatic Control}, 2018.

\bibitem{vaidya2012matrix}
Nitin Vaidya.
\newblock Matrix representation of iterative approximate byzantine consensus in
  directed graphs.
\newblock {\em arXiv preprint arXiv:1203.1888}, 2012.

\bibitem{xie2018generalized}
Cong Xie, Oluwasanmi Koyejo, and Indranil Gupta.
\newblock Generalized {Byzantine}-tolerant sgd.
\newblock {\em arXiv preprint arXiv:1802.10116}, 2018.

\bibitem{yang2017byrdie}
Zhixiong Yang and Waheed~U. Bajwa.
\newblock Byrdie: {Byzantine}-resilient distributed coordinate descent for
  decentralized learning, 2017.

\end{thebibliography}
